\documentclass[10pt, letterpaper]{article}

\usepackage[noadjust]{cite}
\usepackage{amsmath,amssymb}
\usepackage{mathtools}
\usepackage{color,soul}
\usepackage{graphicx}
\usepackage{multirow}
\usepackage{hyperref}
\usepackage{dsfont}
\usepackage{float}
\usepackage{comment}
\usepackage{subfigure}
\usepackage{bm}

\newtheorem{definition}{Definition}

\usepackage[margin=1in]{geometry}
\usepackage{framed}
\usepackage[space]{grffile}
\usepackage[font=footnotesize, skip=6pt]{caption}
\usepackage{tabularx}
\usepackage{gensymb}
\usepackage{scrextend}
\usepackage[ruled,vlined]{algorithm2e}
\usepackage{algorithmic}
\usepackage{booktabs}
\usepackage{subfigure}
\usepackage{indentfirst}
\usepackage{tikz}
\usetikzlibrary{positioning,chains,fit,shapes,calc}
\usetikzlibrary{backgrounds}
\usetikzlibrary{arrows.meta}
\usetikzlibrary{shapes,snakes}

\usepackage{pgfplots}
\usetikzlibrary{positioning,chains,fit,shapes,calc}
\pgfplotsset{compat=1.12}
\title{Stabilizing Fractional Dynamical Networks Suppresses Epileptic Seizures}
\author{
Yaoyue Wang\thanks{Y. Wang (corresponding author) and P. Bogdan are with the Ming Hsieh Electrical and Computer Engineering Department, University of Southern California, USA. {\tt\small yaoyuewa@usc.edu, pbogdan@usc.edu}}, Arian Ashourvan\thanks{A. Ashourvan is with the Psychology Department, University of Kansas, USA. {\tt\small ashourvan@ku.edu}}, Guilherme Ramos\thanks{G. Ramos is with Dept. of Computer Science and Engineering, Instituto Superior Técnico, University of Lisbon, Portugal and Instituto de Telecomunicações, 1049-001 Lisbon, Portugal.
    {\tt\small guilherme.ramos@tecnico.ulisboa.pt}}, Paul Bogdan\footnotemark[1], and Emily Pereira\thanks{Emily Pereira is with the Department of Electrical and Computer Engineering, Texas Tech University, USA. {\tt\small emily.pereira@ttu.edu}}}
\date{}

\begin{document}

\maketitle

\section*{Abstract}
Medically uncontrolled epileptic seizures affect nearly 15 million people worldwide, resulting in enormous economic and psychological burdens. 
Treatment of medically refractory epilepsy is essential for patients to achieve remission, improve psychological functioning, and enhance social and vocational outcomes.
Here, we show a state-of-the-art method that stabilizes fractional dynamical networks modeled from intracranial EEG data, effectively suppressing seizure activity in 34 out of 35 total spontaneous episodes from patients at the University of Pennsylvania and the Mayo Clinic. 
We perform a multi-scale analysis and show that the fractal behavior and stability properties of these data distinguish between four epileptic states: interictal, pre-ictal, ictal, and post-ictal. 
Furthermore, the simulated controlled signals exhibit substantial amplitude reduction ($49\%$ average). These findings highlight the potential of fractional dynamics to characterize seizure-related brain states and demonstrate its capability to suppress epileptic activity.

\noindent\textbf{Keywords:} epilepsy, fractional-order systems, seizure control, intracranial EEG, network dynamics 
\section*{Introduction}
Epilepsy, a disease characterized by unprovoked seizures in the brain, affects more than 50 million people worldwide, and it accounted for \$200 billion in direct medical costs in the U.S. alone between 1993-2019~\cite{menon2025epileptoconomics}. 
Effective and efficient treatment relies on accurately characterizing seizure dynamics to design targeted interventions~\cite{acharyaBrainModelingControl2022}. 

To address these challenges, several questions present themselves, including 
\begin{enumerate}
    \item Can we develop a mathematical framework to distinguish the four epileptic states: interictal, pre-ictal, ictal, and post-ictal? 
    \item Can we leverage this mathematical framework to develop a personalized method to suppress epileptic seizures? 
\end{enumerate}

\begin{figure}[htbp]
\centering\includegraphics[width=\linewidth]{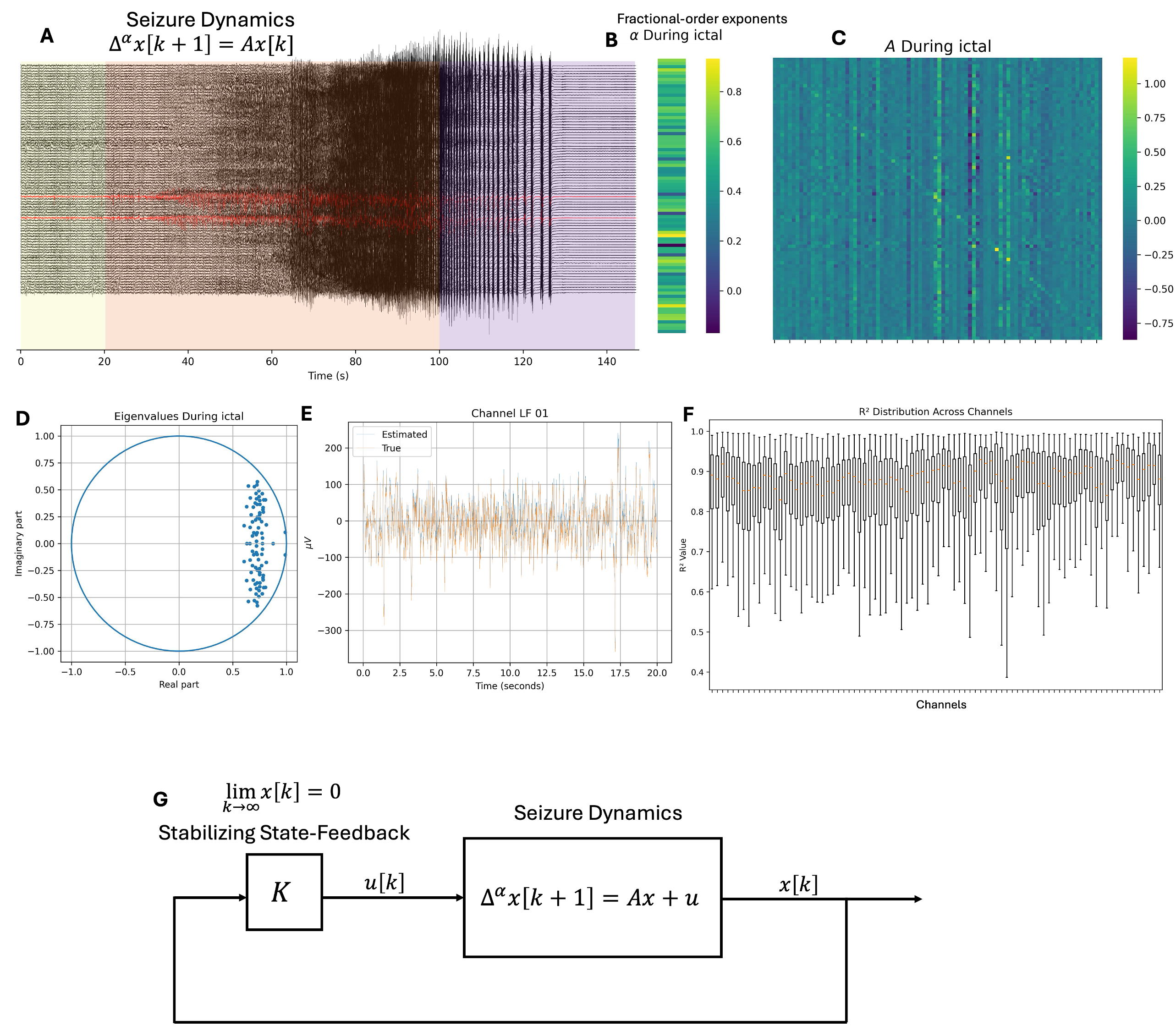}
    \caption{(A) shows an epileptic seizure snapshot recorded using iEEG from patient HUP64, where yellow shading indicates the pre-ictal period, orange shading indicates the ictal period, and purple shading indicates the post-ictal period. (B) shows the fractional-order exponents ($\alpha$) estimated during the beginning of the ictal period. (C) shows the spatial matrix ($A$) estimated during the beginning of the ictal period. (D) shows the eigenvalues of the fractional-order dynamical system during the ictal period. (E) shows a snapshot of the fitted model compared with the true data for a single channel LF01. (F) shows a measure of the best fit of the model ($R^2$) for all channels across the entire seizure snapshot. (G) shows the novel feedback control architecture that we validate on 35 real-world seizure snapshots.}
    \label{fig:figure1}
\end{figure}

An accurate mathematical framework for characterizing seizure dynamics is essential for effective treatment. 
Dynamical network models~\cite{bassett2017network} have emerged as a clear winner, allowing researchers to draw conclusions regarding the brain's topology and function.
Many works have focused on linear time-invariant dynamical networks to model and even control for epileptic activity~\cite{ashourvan2020model, li2017fragility,pequito2017spectral}.
However, recent evidence suggests that the brain exhibits multi-scale dynamics~\cite{presigny2022colloquium,betzel2017multi}, which is not captured by linear time-invariant models~\cite{zhang2012remarks,west2016fractional,shlesinger1993strange}. 
Fractional-order dynamical networks, which originated in physics and economics and quickly found their way into engineering applications~\cite{kilbas2006theory,valerio2013fractional,west2014colloquium,baleanu2012fractional,sabatier2007advances,podlubny1998fractional,petravs2011fractional}, offer a middle ground between linear simplicity and multi-scale complexity.
These networks accurately capture the multi-scale dynamics present in neural signals as well as the spatial relationship between brain regions~\cite{reed2022fractional}. 
Hence, fractional-order dynamical networks outperform traditional integer-order models in accurately representing neural data~\cite{baleanu2011fractional,west2016networks,moon2008chaotic,lundstrom2008fractional,werner2010fractals,thurner2003scaling,teich1997fractal}. 

\begin{figure}[t!]
    \centering
    \includegraphics[width=\textwidth]{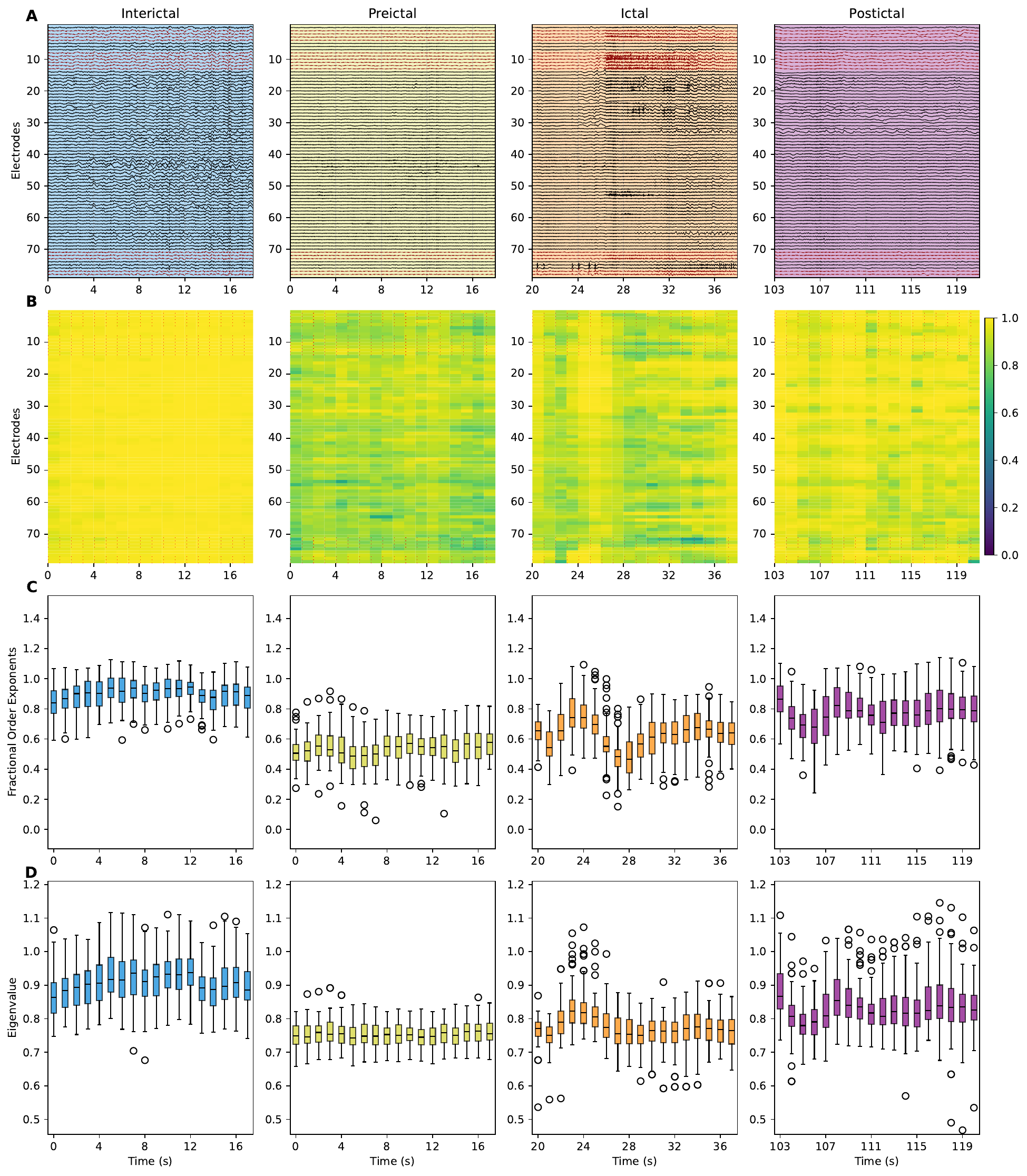}
    \caption{Patient HUP68, Seizure 5. Columns: epileptic brain states (interictal, pre-ictal, ictal, post-ictal). Rows show data across all electrodes: (A) filtered iEEG signals from 79 electrodes with seizure onset zone (SOZ) electrodes in red, (B) heatmap of R$^2$ values across all electrodes for all 3-second sliding time windows, (C) boxplots of estimated fractional-order exponents, and (D) boxplots of eigenvalues.}
    \label{fig:HUP68_4x4_block_5}
\end{figure}

In this work, we provide a fractional-order dynamical framework to comprehensively characterize seizure dynamics across all epileptic states. 
Figure~\ref{fig:HUP68_4x4_block_5} illustrates this process, which analyzes fractional-order exponents ($\alpha$) and system eigenvalues ($\lambda$) to assess fractional dynamics and network stability properties across interictal, pre-ictal, ictal, and post-ictal brain states.
By leveraging our fractional-order dynamical network mathematical framework, we validate a novel stabilizing control strategy on 35 seizures from 10 patients. 

Our approach provides a comprehensive mathematical framework for understanding seizure evolution and for designing effective personalized seizure control. Through explicitly modeling multi-scale and stability properties in epileptic dynamics, we offer insights into seizure mechanisms and an approach to effectively suppress seizures.

\section*{Results}

\subsection*{Fractional-Order Dynamical Networks Accurately Fit Epileptic Data Over the Majority of Segments Across All Seizure States}

We assessed the goodness of model fit by computing the $R^2$ value for 2656 electrodes in all four epileptic brain states (interictal, pre-ictal, ictal, and post-ictal) for each patient. Segments that had an $R^2<0.5$ for 40\% or more of their electrodes were removed from analysis to avoid unreliable results due to poor model fit (these segments were analyzed separately -- see Supplementary Material). Based on our $R^2$ criterion for model validation, we excluded 14 of 140 segments (10\%) from analysis (see Table~1 Supplementary Material), leaving 126 valid segments. 

The omitted segments consisted of 8 post-ictal segments (HUP64: 1, HUP68: 3, HUP78: 1, HUP86: 2, MAYO016: 1), all 5 interictal segments from HUP78, and 1 pre-ictal segment from HUP78. The excluded post-ictal segments exhibited significantly smaller signal amplitudes (139.1 $\pm$ 177.0 µV) compared to included post-ictal segments (457.8 $\pm$ 928.7 µV), while HUP78's interictal segments showed abnormally high amplitudes (238.5 $\pm$ 176.3 µV) relative to other patients' interictal segments (153.4 $\pm$ 126.7 µV). 

\begin{figure}[th!]
    \centering
    \includegraphics[width=\textwidth]{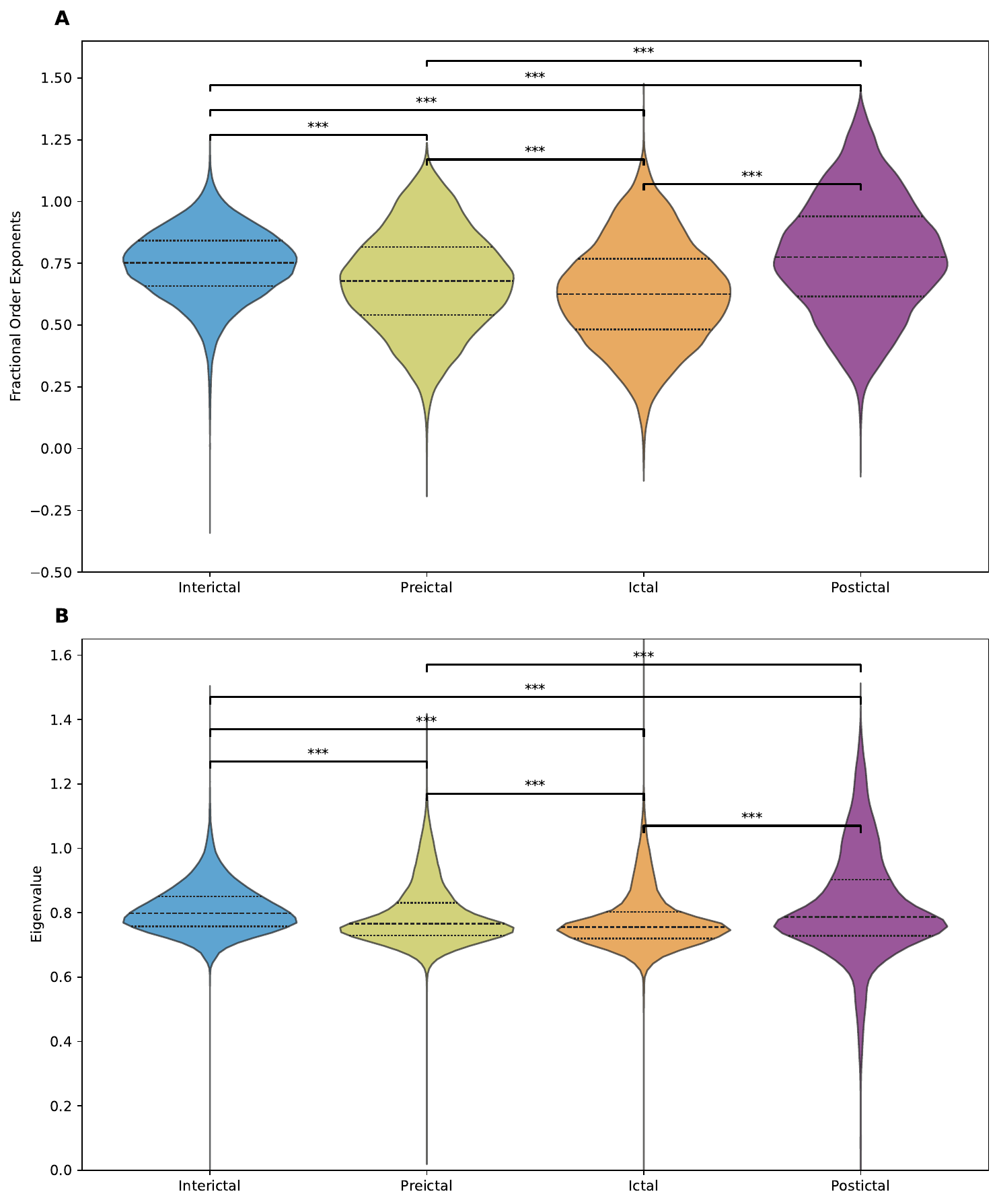}
    \caption{Violin plots of multi-scale and stability properties across all electrodes, for each epileptic brain state. 
    Data are only pooled from segments where at least $60\%$ of electrodes have an $R^2\geq0.5$ across all time windows within each segment. *** indicates statistically significant differences between all pairs of epileptic brain states ($p<.001$, Kolmogorov-Smirnov test). 
    (A) Fractional-order exponents ($\alpha$) characterizing multi-scale properties of iEEG signals. 
    (B) Eigenvalues of the system $A_0$ matrix characterizing stability properties of fractional-order dynamical models.}
    \label{fig:alpha_eigen_summary_all}
\end{figure}

\subsection*{Changes Across Epileptic Brain States Evident in Multi-scale and Stability Properties with Consistent Observations across Patients}
After establishing an appropriate model fit, we illustrate the temporal evolution of the raw intracranial EEG (iEEG) data, goodness of fit, fractional-order exponents ($\alpha$), and eigenvalues during a complete seizure cycle of patient HUP68's 5th seizure (Figure~\ref{fig:HUP68_4x4_block_5}). 
Panel A in Figure~\ref{fig:HUP68_4x4_block_5} shows the iEEG recordings from 79 electrodes, with seizure onset zone (SOZ) electrodes highlighted in red. 
Visual inspection of iEEG data reveals distinct activity patterns across segments, with the ictal period showing high-amplitude, rhythmic discharges that are particularly prominent in the SOZ electrodes. 
In contrast, the iEEG data during the pre-ictal period exhibits a smaller signal amplitude across all electrodes. 
Similarly, the iEEG data from SOZ electrodes during the interictal period have smaller amplitudes compared to non-SOZ electrodes. 

The heatmap in panel B in Figure~\ref{fig:HUP68_4x4_block_5} shows $R^2$ values for each electrode and time window, illustrating the goodness of fit for the fractional-order dynamical networks across all epileptic brain states. Like the iEEG data, each SOZ electrode has a border of red dots. We observe large $R^2$ values across all segments, indicating good model performance across all segments. 
 
Panel C in Figure~\ref{fig:HUP68_4x4_block_5} shows the fractional-order exponents ($\alpha$) across all four states. Interictal fractional-order exponents are stable and tightly clustered around 0.8-1.0, indicating low memory dependence and decreased multi-scale properties. 
Pre-ictal fractional-order exponents are smaller than fractional-order exponents during interictal and have a larger spread in distribution, indicating a higher memory dependence and increased multi-scale properties. 
Fractional-order exponents during the ictal period are slightly larger than the fractional-order exponents during the pre-ictal period and have the most variability overall, indicating potentially chaotic behavior and inconsistent multi-scale properties. 
Post-ictal fractional-order exponents are slightly larger than fractional-order exponents during ictal, with increased variability, evident in the wider IQR and outliers, indicating a shift in memory dependence and multi-scale properties.

Panel D in Figure~\ref{fig:HUP68_4x4_block_5} shows the eigenvalues across all epileptic brain states.
Eigenvalues during interictal have larger variance, with some values exceeding 1, indicating unstable or potentially marginally stable behavior. 
Eigenvalue distributions during pre-ictal tighten, with all values less than 1, reflecting a stable system. 
Eigenvalues during ictal have slightly larger values and a wider spread than eigenvalues during pre-ictal, suggesting that the system is moving closer towards instability. 
Finally, eigenvalues during post-ictal are larger, have a greater IQR, and have significantly more outliers than eigenvalues during ictal, confirming post-seizure instability. Interestingly, the interictal state has the largest eigenvalues. 

\begin{figure}[t!]
    \centering
    \includegraphics[width=\textwidth]{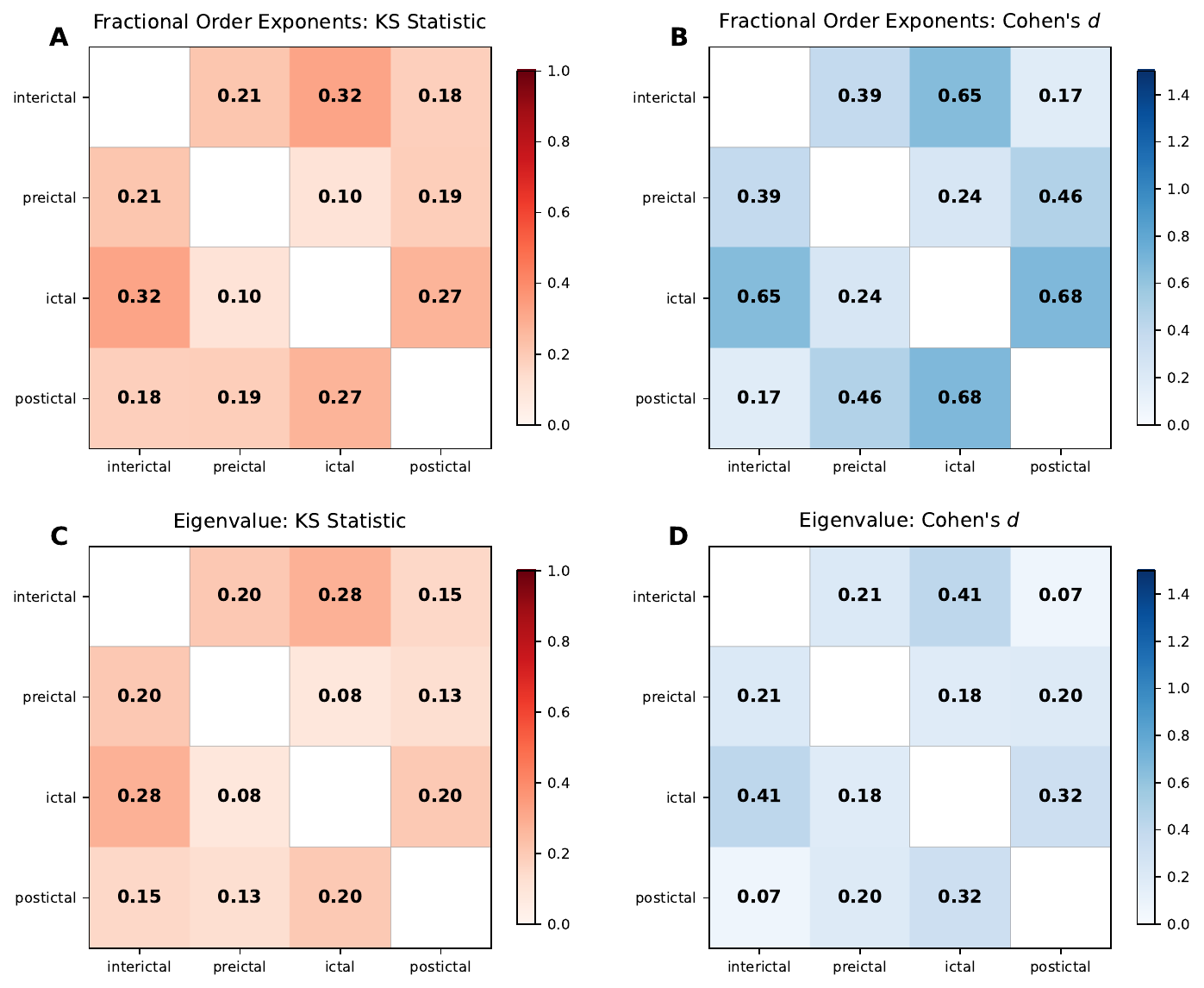}
    \caption{Population-pooled pairwise comparisons across epileptic brain states. 
    (A) KS statistic for fractional-order exponents ($\alpha$) comparisons. 
    (B) Cohen's $d$ effect sizes for fractional-order exponents ($\alpha$) comparisons. 
    (C) KS statistic for eigenvalue comparisons. 
    (D) Cohen's $d$ effect sizes for eigenvalue comparisons. 
    All data are pooled across patients within each brain state before computing statistics. 
    All comparisons reached statistical significance ($p<0.001$). 
    Effect sizes range from small to medium, with fractional-order exponents ($\alpha$) showing larger differences than eigenvalues and ictal showing the largest separation from other states for both metrics.} 
    \label{fig:heatmap_pooled}
\end{figure}

\begin{figure}[t!]
    \centering
    \includegraphics[width=\textwidth]{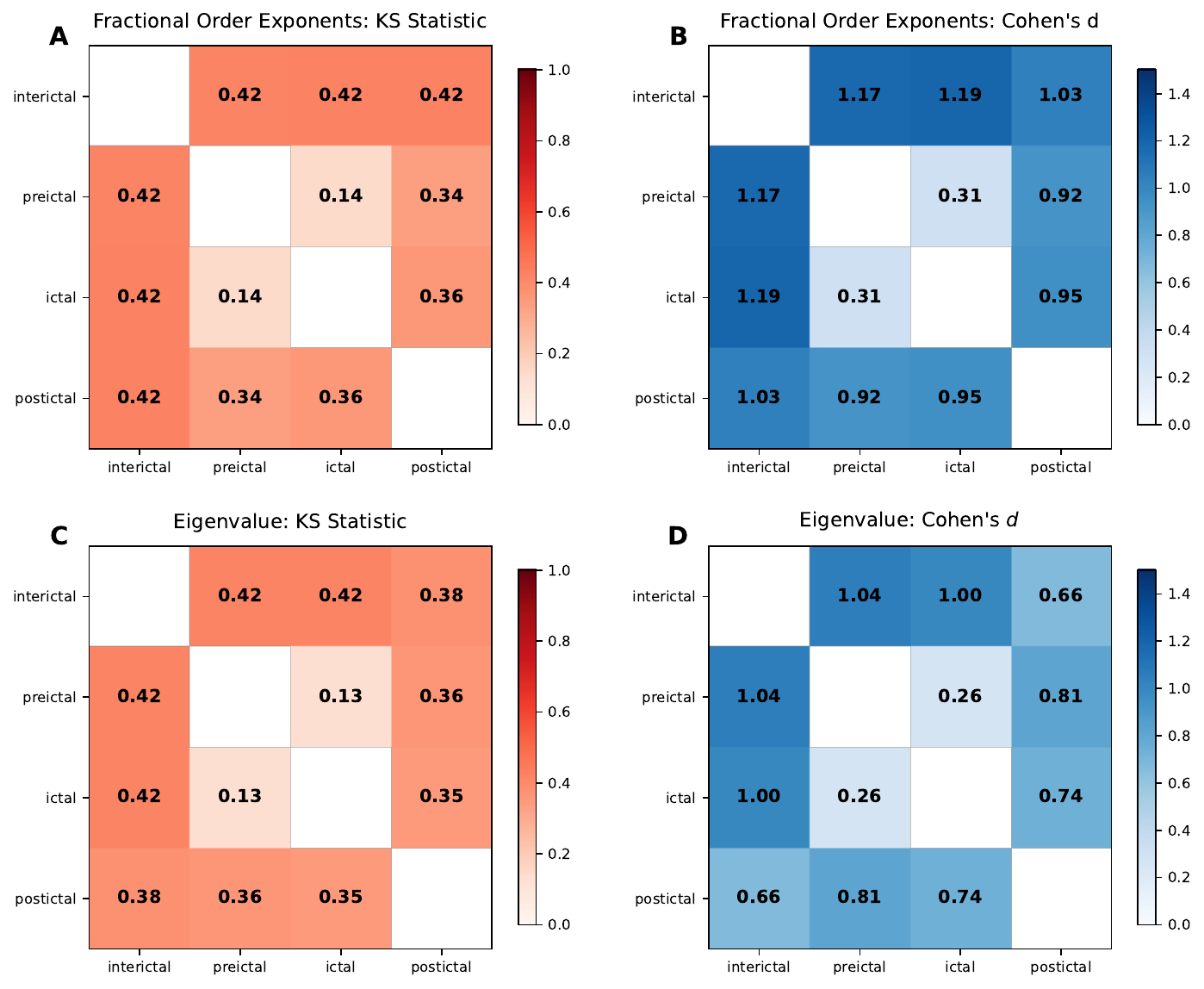}
    \caption{Within-patient pairwise comparisons across epileptic brain states, averaged across patients. 
    (A) KS statistic for fractional-order exponents ($\alpha$) comparisons. 
    (B) Cohen's $d$ effect sizes for fractional-order exponents ($\alpha$) comparisons. 
    (C) KS statistic for eigenvalue comparisons. 
    (D) Cohen's $d$ effect sizes for eigenvalue comparisons. 
    Statistics are computed within each patient and then averaged across patients to account for inter-patient variability. 
    Effect sizes are substantially larger than population-pooled analysis, with large effect sizes (Cohen's $d \geq 0.8$) for fractional-order exponents ($\alpha$) and medium-to-large effect sizes for eigenvalues. 
    Interictal shows the most consistent separation from all other states across both metrics.}
    \label{fig:heatmap_within_patient}
\end{figure}

\subsubsection*{Multi-scale Property Shows Distinct Shifts Across Epileptic Brain States}
 
Figure~\ref{fig:alpha_eigen_summary_all} A shows the fractional-order exponents ($\alpha$) distributions for each epileptic brain state across all patients and electrodes. 
Median fractional-order exponents decreased from interictal 0.75 (IQR: 0.66-0.84) through pre-ictal 0.68 (IQR: 0.54-0.82) and ictal 0.63 (IQR: 0.48-0.77), and then increased during post-ictal 0.78 (IQR: 0.62-0.94). 
Interictal and post-ictal segments exhibit fractional-order exponents with extended tails, with fractional-order exponents during skewing toward smaller values and fractional-order exponents during post-ictal showing a wider spread in both directions. 
In contrast, fractional-order exponents ($\alpha$) during pre-ictal and ictal segments have more uniform distributions, though fractional-order exponents during ictal segments show larger outliers. 

Several consistent patterns emerged across patients (see Supplementary Material Table~2). 
Fractional-order exponents ($\alpha$) during interictal cluster around 0.70-0.80 for most patients, while fractional-order exponents during pre-ictal and ictal segments remained relatively similar within individual patients. 
Fractional-order exponents during post-ictal periods exhibited increased variability across most patients, with the exception of HUP72 and MAYO010. 
Six out of ten patients (HUP64, HUP68, HUP70, HUP86, MAYO011, MAYO016) showed decreased median fractional-order exponents ($\alpha$) from interictal to ictal, consistent with the group-level trajectory. Additionally, MAYO patients generally showed larger post-ictal fractional-order exponents ($\alpha$) compared to the HUP patients. 

Despite these overall trends, patient-level fractional-order exponents ($\alpha$) distributions varied substantially (see Supplementary Material Figure~2 and Table~2). 
Patients MAYO011 and MAYO016 showed large fractional-order exponents during post-ictal (medians = 1.02 and 0.90, respectively), whereas MAYO010 and HUP70 had consistent fractional-order exponents across all segment types. 
HUP72 exhibited distinctly large fractional-order exponents during pre-ictal and ictal segments (medians = 0.99 and 0.97). 

\subsubsection*{Network Stability Properties Reveal Distinct Patterns Across Epileptic Brain States}

Figure~\ref{fig:alpha_eigen_summary_all} B shows the eigenvalue distributions across all patients and electrodes.
The eigenvalues showed less variability compared to fractional-order exponents ($\alpha$). 
Eigenvalues during interictal showed the tightest distribution across all patients (median = 0.80, IQR: 0.76-0.85) while eigenvalues during pre-ictal, ictal, and post-ictal segments exhibited similar median values (0.73, 0.72, and 0.72, respectively). 
However, eigenvalues during post-ictal showed increased variability (IQR: 0.73-0.90) compared to all other segments (pre-ictal IQR: 0.73-0.83, ictal IQR:0.72-0.80). 
All epileptic brain states showed eigenvalue distributions with extended tails and numerous outliers, far greater than those of fractional-order exponents ($\alpha$), especially during post-ictal. 

Patient-level eigenvalue distributions revealed patterns consistent with group-level findings (see Supplementary Material Figure~3 and Supplementary Material Table~3). 
Eigenvalues during interictal were relatively stable across most patients (median $\approx$0.75), while eigenvalues during pre-ictal and ictal remained consistent within individual patients, mirroring the pattern observed for fractional-order exponents ($\alpha$). 
Eigenvalues during post-ictal periods exhibited notably increased spread, evident in the broadened violin distributions in all patients except HUP70, HUP72, and HUP78. 
HUP70 showed consistent eigenvalues across all brain states (consistent with its fractional-order exponents ($\alpha$) patterns), while HUP72 showed smaller eigenvalues during post-ictal (median = 0.62, IQR: 0.54–0.75).
Across all patients, we observed stable eigenvalues during interictal periods, with minimal variation between eigenvalues during pre-ictal and ictal. 
Eigenvalues during post-ictal periods are characterized by increased variability and have larger fractional-order exponents ($\alpha$) and eigenvalues.

\subsection*{Stronger Effect Sizes within Patients than at Population Level}

We performed pairwise comparisons of fractional-order exponents ($\alpha$) and eigenvalues of all epileptic brain states using population-pooled and within-patient analysis to assess the separation between epileptic brain states. 

Our population pooled analysis (see Figure~\ref{fig:heatmap_pooled}) using the Kolmogorov-Smirnov statistical test showed significance between all epileptic brain states ($p <.001$), but with small to medium effect sizes. 
We observed the largest Cohen's $d$ values when comparing fractional-order exponents during ictal and post-ictal (0.68) as well as during ictal and interictal (0.65) -- see Figure~\ref{fig:heatmap_pooled} B.
When comparing eigenvalues during ictal and post-ictal, we observed Cohen's $d$ value of 0.41, and for eigenvalues during ictal and interictal, we reported a Cohen's $d$ value of 0.32 -- see Figure~\ref{fig:heatmap_pooled} D. 
Kolmogorov-Smirnov statistic values shown Figure~\ref{fig:heatmap_pooled} A and C were similar for both fractional-order exponents and eigenvalues. 

Our within-patient analysis (see Figure~\ref{fig:heatmap_within_patient}) revealed substantially larger effect sizes compared with our population pooled analysis. 
For comparing fractional-order exponents ($\alpha$) shown in Figure~\ref{fig:heatmap_within_patient} B, we observed that Cohen's $d$ value exceeded $0.9$ for all comparisons except for fractional-order exponents during pre-ictal vs. ictal. 
We observed the largest effect sizes for fractional-order exponents during interictal vs. ictal ($d = 1.19$) and fractional-order exponents during interictal vs. pre-ictal ($d = 1.17$). 
When comparing eigenvalues during epileptic brain states using the Kolmogorov-Smirnov statistical test shown in Figure~\ref{fig:heatmap_within_patient} D, we observed medium to large effect sizes. 
Comparing eigenvalues during interictal vs. pre-ictal showed Cohen's $d = 1.04$, and comparing eigenvalues during interictal vs. ictal showed Cohen's $d = 1.00$, which showed the strongest separation (see Figure~\ref{fig:heatmap_within_patient} D). 
Kolmogorov-Smirnov statistics shown in Figure~\ref{fig:heatmap_within_patient} A showed consistently large Cohen's $d$ values for fractional-order exponents during interictal in comparison with all other epileptic brain states (0.42 for all three comparisons), indicating robust distributional separation within individual patients.
Kolmogorov-Smirnov statistics shown in Figure~\ref{fig:heatmap_within_patient} D showed consistent Cohen's $d$ values for eigenvalues as observed in Figure~\ref{fig:heatmap_within_patient} A for fractional-order exponents. 

Adjacent epileptic brain states (i.e., interictal vs. pre-ictal, pre-ictal vs. ictal, and ictal vs. post-ictal) showed smaller effect sizes in both individual and population analyses. 
Both fractional-order exponents and eigenvalue comparisons during pre-ictal vs. ictal yielded the smallest effect size with $d = 0.31$ for fractional-order exponents and $d = 0.26$ for eigenvalues. 
On the other hand, both fractional-order exponents and eigenvalue comparisons during ictal vs. post-ictal showed larger effect sizes with $d = 0.95$ for fractional-order exponents and $d = 0.74$ for eigenvalues.

The contrast between population-pooled and within-patient effect sizes suggests substantial inter-patient heterogeneity in baseline epileptic dynamics. 
While individual patients show large, consistent differences between epileptic brain states, pooling across patients reduces these effects due to varying baseline values and seizure characteristics across the cohort.

\begin{figure}[t!]
    \centering
    \includegraphics[width=\textwidth]{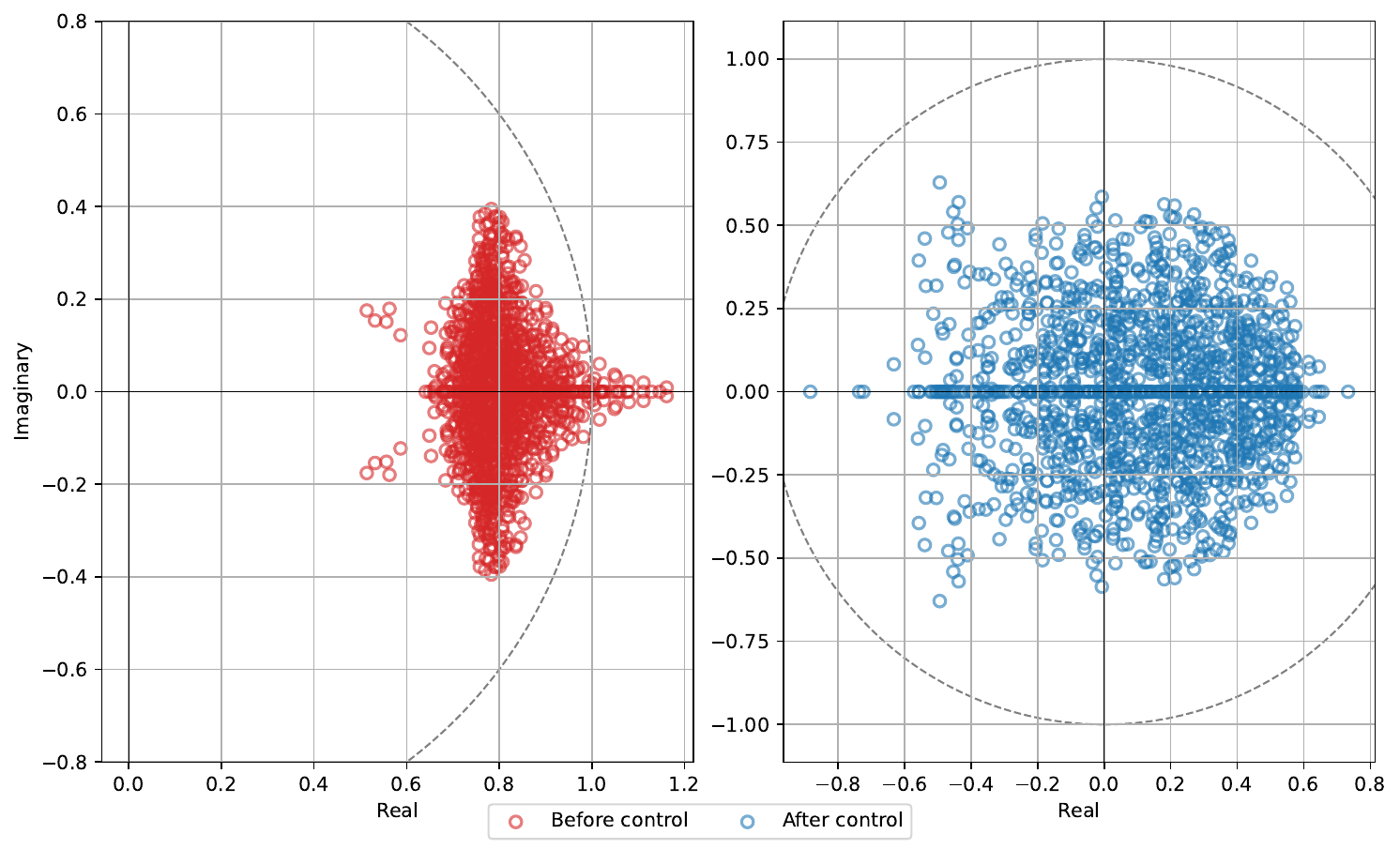}
    \caption{Patient HUP68 Seizure 3. Red shows eigenvalues of the system matrix $A_0$ on the complex plane during seizure onset before applying stabilizing control. The fractional-order dynamical network is unstable, as at least one eigenvalue is outside of the unit circle on the complex plane. Blue shows eigenvalues of the simulated system matrix $A_0$ during seizure onset after applying stabilizing control. All eigenvalues are stabilized and lie within the unit circle on the complex plane, restoring system stability.}
    \label{fig:HUP68_ctrl_block_3}
\end{figure}

\begin{figure}[t!]
    \centering
    \includegraphics[width=\textwidth]{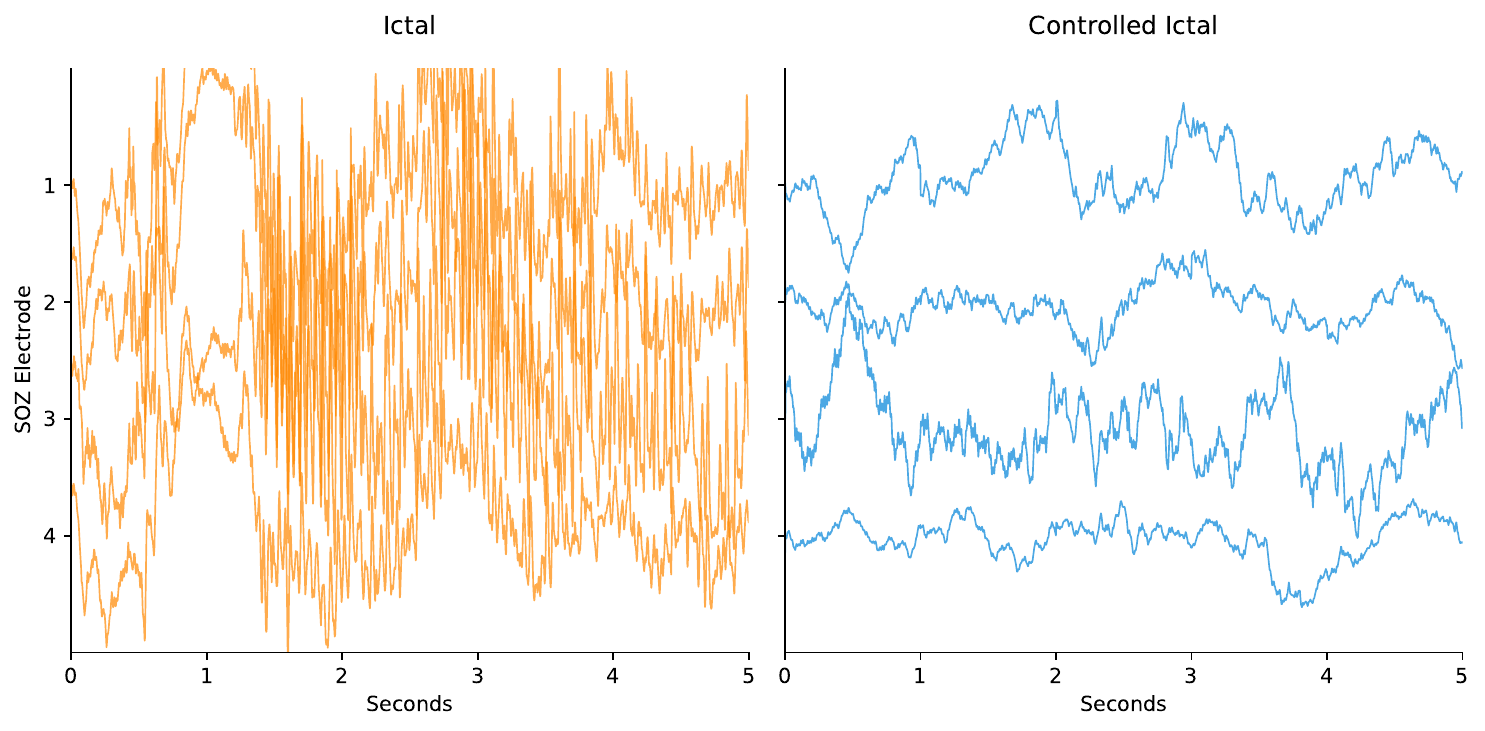}
    \caption{Patient HUP68 seizure 3. (A) iEEG seizure onset zone (SOZ) signals from 4 electrodes during seizure onset before applying stabilizing control (mean amplitude: 159.03~$\mu$V). (B) Simulated controlled SOZ signals from the same electrodes after applying stabilizing control (mean amplitude: 60.54~$\mu$V). The control reduces the signal amplitude by 61.9\%.}
    \label{fig:HUP68_3_raw_vs_ctrl}
\end{figure}

\begin{figure}[t!]
    \centering
    \includegraphics[width=\textwidth]{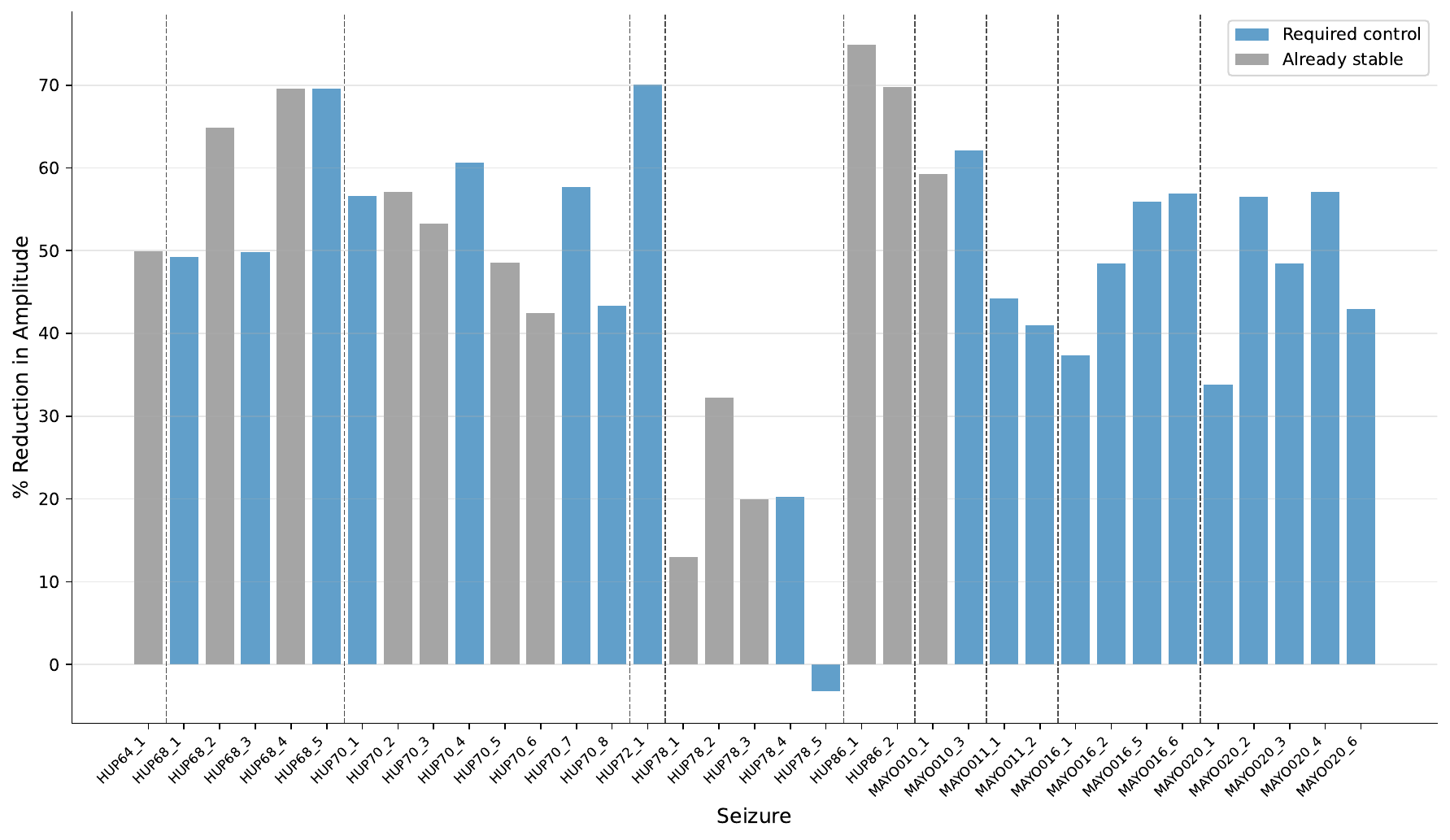}
        \caption{Percentage amplitude reduction for each seizure. Gray bars represent seizures that are already stable (maximum eigenvalues $\leq 1$), while blue bars indicate seizures that require stabilization. Vertical dashed lines separate patients. Only 1 seizure increased in amplitude after control. Control reduced amplitude by an average of $48.96\% \pm 16.94\%$.}
    \label{fig:amp_reduction}
\end{figure}

\subsection*{Control Stabilizes Most Seizures with Failures Linked to Ill-Conditioned Optimization}

Our stabilizing controller was applied at seizure onset for all ictal segments. 
Figure~\ref{fig:HUP68_ctrl_block_3} illustrates the effectiveness of our control strategy by comparing eigenvalues in the complex plane before and after applying control in Seizure 3 of HUP68. 
The uncontrolled eigenvalues (red) lie near or beyond the stability boundary, while the controlled eigenvalues (blue) are well within the unit circle, satisfying the stability definition (see Methods Definition~\ref{def:stability}). 
Figure~\ref{fig:HUP68_3_raw_vs_ctrl} shows the simulated controlled iEEG data compared to the original iEEG data from the seizure onset zone (SOZ) electrodes for Seizure 3 of HUP68. 
We see that the controlled iEEG signals have substantially smaller amplitudes compared to the uncontrolled ictal iEEG signals. 
Figure~\ref{fig:amp_reduction} shows the percent reduction in amplitude for all electrodes in all 35 seizures. 
We observed an average reduction of $48.96\% \pm 16.94\%$ in signal amplitude. 
There was little difference in amplitude reduction between fractional-order systems that were already stable during ictal and unstable fractional-order systems requiring control ($50.38\% \pm 18.41\%$ and $48.12\% \pm 15.95\%$, respectively). 
 
During the time of seizure onset, 22 out of 35 seizures had unstable fractional-order dynamical networks ($\exists \lambda: |\lambda|>1$), while 13 seizures had stable fractional-order dynamical networks ($\max|\lambda| <1$). 
After applying control, 27 fractional-order dynamical networks during ictal (77\%) satisfied the stability criterion (max$|\lambda| <1$), representing successful stabilization of 17 out of 22 (77\%) initially unstable fractional-order dynamical networks during ictal. 
Eight seizures failed to stabilize: HUP68 seizure 5, HUP78 seizures 1-5, MAYO016 seizure 2, and MAYO020 seizure 6.
To investigate the cause of these failures, we analyzed the numerical properties of the optimized matrices $L$ and $P$. 
Failed cases exhibited significantly larger condition numbers (median $\kappa(L)$: 1382 vs 28 for successful cases), indicating severe ill-conditioning of the optimization problem. 
Failed cases also required substantially larger control corrections (median $\|L\|_F$: 45.4 vs 15.5) and $P$ matrices (median $\|P\|_F$: 51.7 vs 23.0).

\section*{Discussion}

\subsection*{Summary}
We applied fractional-order dynamical systems to model brain network dynamics across seizure progression in 10 patients with drug-resistant epilepsy, examining how stability (eigenvalues) and multi-scale (fractional-order exponents) properties differ across interictal, pre-ictal, ictal, and post-ictal epileptic brain states. 
Our key contributions included achieving good model fit across all epileptic brain states with interpretable properties that revealed distinct patterns and a stabilizing control framework that satisfied the eigenvalue stability criterion in 77\% of unstable ictal networks and reduced seizure signal amplitude by an average of~49\%. These findings provide evidence that multi-scale and stability dynamics evolve distinctly across seizure progression, with inter-patient variability potentially reflecting diverse epilepsy etiologies. 
The integration of fractional-order modeling with control theory provides a foundation for developing personalized, model-based neurostimulation strategies targeting seizure suppression.

\subsection*{Fractional-Order Dynamical Network Models are Suitable to Represent Epileptic Brain States}
Fractional-order dynamical networks achieve $R^2\geq0.5$ for 60\% of all electrodes in 126 of 140 segments (90\%), demonstrating that fractional-order systems effectively capture iEEG dynamics across diverse seizure states in most patients. 
The 14 segments where a good fit could not be obtained likely reflect limitations in our current estimation framework when confronted with extreme signal characteristics, rather than fundamental flaws in modeling the data with fractional-order dynamics. 
Our model validation revealed that the majority of rejected data occurred during the post-ictal segments, which had significantly smaller signal amplitudes (139.1$\pm$177.0 µV) compared to successfully modeled post-ictal segments (457.8$\pm$928.7 µV). 
This pattern may reflect post-ictal suppression, where the brain enters a recovery state with markedly lower electrical activity~\cite{pottkamper2020postictal}. 
This finding suggests that fractional-order dynamical network modeling is most informative when applied to periods with active neural dynamics rather than suppressed brain states. 
Patient HUP78 presented the opposite pattern, as all five of its interictal segments were excluded due to abnormally high amplitudes (238.5$\pm$176.3 µV). 
HUP78 developed epilepsy from a traumatic brain injury, highlighting how individual patient characteristics and epilepsy etiology may influence the applicability of network-based modeling approaches.

\subsection*{Consistent Multi-scale and Stability Properties Across Patients Suggest Reproducible Brain Network Activity}


Our analysis of fractional-order exponents and eigenvalues across seizure states reveals that epileptic brain networks exhibit consistent, multi-scale memory dynamics and structured stability properties. Fractional-order exponents quantify history-dependence in neural dynamics, a property that integer-order models cannot capture. These fractional-order exponents reveal how strongly current brain states depend on past activity across multiple timescales, with lower values indicating stronger memory dependence and higher values indicating weaker memory dependence, and values approaching 1 reducing to the integer-order case. Across patients, we observed a reproducible pattern across epileptic brain states.
Interictal fractional-order exponents indicate low memory dependence and a decrease in multi-scale properties. 
Pre-ictal fractional-order exponents reflect a higher memory dependence and an increase in multi-scale properties. Fractional-order exponents during the ictal period indicate potentially chaotic behavior and inconsistent multi-scale properties. On the other hand, post-ictal fractional-order exponents indicate a shift in memory dependence and multi-scale properties, without returning to interictal levels.
Similarly, eigenvalues during interictal indicate unstable or potentially marginally stable behavior. 
Pre-ictal eigenvalue distributions generally reflect stabilized behavior, while the ictal eigenvalues suggest that the brain is moving closer towards instability. 
Finally, eigenvalues during post-ictal indicate instability, suggesting disrupted or incomplete recovery following seizure termination.

In Figure~\ref{fig:alpha_eigen_summary_all} A, the interictal segments exhibit the weakest memory dependence (highest $\alpha$), while pre-ictal and ictal periods show strong memory effects. Post-ictal fractional-order exponents indicate that these segments are much more chaotic and do not return to interictal levels. This trend is mirrored in Figure~\ref{fig:alpha_eigen_summary_all} B with the eigenvalues and stability. The consistency in fractional-order exponents ($\alpha$) (median 0.75, IQR: 0.66-0.84) and eigenvalues (median 0.80, IQR: 0.76-0.85) across interictal periods reveals that epileptic brain networks maintain consistent dynamics between seizures. 
These uniform fractional-order exponent distributions indicate that interictal epileptic dynamics are fundamentally non-Markovian, with neural activity substantially coupled to its temporal history~\cite{west2016fractional, lundstrom2008fractional}. We consistently observed eigenvalues greater than 1 during interictal periods, indicating unstable dynamics. 
Our within-patient analysis revealed that eigenvalues during interictal have the strongest separation from other epileptic brain states, with large effect sizes (Cohen's $d>1$) for all comparisons except for eigenvalues during interictal vs post-ictal.

These patterns and strong effect sizes highlight the structured nature of seizure dynamics, aligning with prior studies that have used EEG data to track physiological changes~\cite{schwamb2024data}. Our findings are also consistent with the theory of critical slowing, which has been proposed as a biomarker for seizure susceptibility~\cite{maturana2020critical}. 

Seizure onset and termination have been modeled as critical transitions in dynamical systems~\cite{kramer2012human,scheffer2009early,maturana2020critical,moosavi2022critical,touboul2011neural}, and recent work has explored seizure dynamics across spatial scales~\cite{kramer2012human,martinet2017human}. Other studies have identified self-similar properties in EEG seizure data using multivariate eigen-wavelet analysis~\cite{lucas2023epileptic}, generalized Hurst exponents to distinguish healthy from epileptic activity~\cite{lahmiri2018generalized}, and fractal features for machine learning classification in a clinical context~\cite{francca2018fractal}. 
Our work echoes the conclusions of these previous studies by explicitly modeling scale-dependent memory and stability, offering a unified framework to characterize and track multiscale seizure dynamics across time.

\subsection*{Multi-scale and Stability Properties Uncover Insights into Seizure Onset, Progression, and Recovery}


The shift from interictal to ictal periods showed a decrease in fractional-order exponents ($\alpha$) (0.75 to 0.63) and eigenvalues (0.80 to 0.72). Lower fractional-order exponents ($\alpha$) indicates stronger history-dependence of neural activity, which suggests that seizures are self-sustaining events, where the brain cannot escape from pathological activity patterns~\cite{jiruska2013synchronization}. Similarly, eigenvalues also decreased in pre-ictal and ictal periods, with somewhat tightened distributions, suggesting a modest shift toward a more robust, stable region during seizures. While this observation may initially appear to contradict recent literature that hypothesizes that seizures can be mathematically represented as an instability in the brain~\cite{sritharan2014fragility,qin2024analytical,touboul2011neural,jirsa2014nature,le1997unstable}, we note that asymptotic stability (assessed in this work) may not be suitable for characterizing seizure onset since seizure events have finite time horizons. Notably, the stability of a fractional-order dynamical network depends on both its fractional-order exponents ($\alpha$) and its spatial matrix ($A$). In this work, we observe that fractional-order exponents evolve together with eigenvalues across epileptic brain states. 

Importantly, both fractional-order exponents ($\alpha$) and eigenvalues show small differences between \mbox{pre-ictal} and ictal segments. The similarity between \mbox{pre-ictal} and ictal suggests that the pathological network states characteristic of seizures are already established before clinical seizure onset, consistent with previous findings that seizures do not arise instantaneously but emerge from gradual network-level transitions~\cite{litt2001epileptic, mormann2007seizure}. Furthermore, the pre-ictal period may represent an already altered brain state that is primed for seizure generation rather than a simple precursor to an abrupt change~\cite{jiruska2013synchronization, khambhatiDynamicNetworkDrivers2015}.

Post-ictal fractional-order exponents ($\alpha$) and eigenvalues did not return to interictal baseline values, revealing substantial heterogeneity in recovery dynamics. Some patients exhibited both fractional-order exponents and eigenvalues exceeding interictal baselines, suggesting hyperexcitable rebound, while others showed sustained suppression. The variability within post-ictal periods (wider IQR and numerous outliers for both metrics) indicates unstable network dynamics during recovery, where the brain may cycle through multiple states before re-establishing a baseline state. This finding indicates that post-seizure recovery is not a uniform process but rather reflects patient-specific network reorganization mechanisms~\cite{pottkamper2020postictal, Lamberts2013, Bauer2017}.

Within-patient analysis revealed substantially larger Kolmogorov-Smirnov test statistics and effect sizes compared to the population-pooled statistics. The strong within-patient separation between interictal and seizure-related states suggests that fractional-order exponents could serve as patient-specific biomarkers for predicting seizures, despite being limited on the population level. 
Additionally, since fractional-order exponents during pre-ictal and ictal states showed minimal differences, detecting the transition from interictal to pre-ictal may provide sufficient warning for impending seizures without requiring the exact time of seizure onset. This is further supported by the observation that interictal to ictal, and interictal to pre-ictal comparisons exhibited similarly strong separability, as shown in Figure~\ref{fig:heatmap_within_patient}.

The pattern of effect sizes reveals an asymmetry in seizure dynamics. The sharp contrast between ictal and post-ictal states ($d = 1.00$ for fractional-order exponents and $0.79$ for eigenvalues) compared to the minimal difference between pre-ictal and ictal states ($d = 0.28$ for fractional-order exponents and $0.25$ for eigenvalues) may indicate that seizure termination triggers more dramatic network reorganization than seizure initiation. 
This finding aligns with recent modeling work showing that seizure onset and spread can emerge as critical transitions in brain networks driven by changes in excitability and connectivity~\cite{moosavi2022critical}. Furthermore, our finding that fractional-order exponents and eigenvalues separate interictal and pre-ictal states is supported by prior work highlighting the importance of identifying network-level changes as patient-specific biomarkers for seizure prediction~\cite{assi2017towards}.

\subsection*{Stabilizing Control Shows Promise to Suppress Epileptic Activity}
To our knowledge, our work represents one of the most comprehensive demonstrations of fractional-order control for suppressing seizure dynamics applied to real patient iEEG data. Our stabilizing control framework achieved seizure suppression in 34/35 seizures, successfully stabilizing 77\% of initially unstable seizures and reducing seizure amplitude by approximately 49\%  across all electrodes. Notably, our control strategy produced a similar average amplitude reduction in both initially stable and unstable seizures, showcasing effective suppression of ictal activity regardless of mathematical stability status. These findings suggest that fractional-order control strategies can generalize across diverse patient pools.

In contrast, previous work has focused on developing state-feedback stabilization strategies for linear time-invariant models of physiological networks~\cite{ehrens2015closed}. Another example is a generalized pole placement algorithm for linear time-invariant dynamics~\cite{ashourvan2020model}. The main issue with linear-time invariant models is that they do not capture the measurement dependence inherent in physiological networks~\cite{reed2022fractional}.

Recent work has focused on designing linear networked models~\cite{nobili2023vibrational,qin2025vibrational,allibhoy2022optimal}. For example, the work in~\cite{nobili2023vibrational} derives graph-theoretic conditions for structural vibrational stabilizability under which linear networked models can always be stabilized, and the work in~\cite{qin2025vibrational} provides a method to design the vibrational inputs to stabilize the linear network. The work in~\cite{allibhoy2022optimal} studies interconnected excitatory-inhibitory pairs with linear threshold dynamics, and presents strategies to design networks with desired robustness properties. These methods may offer theoretical guarantees under ideal conditions, but are not easily estimated from data and may not accommodate the heterogeneity observed across patients. 

Open-loop vibrational control strategies have been proposed to stabilize networks of nonlinear oscillators~\cite{qin2022vibrational,menara2022functional,qin2023vibrational}. For example, recent work shows that vibrational control synchronizes clusters of nonlinear oscillators in a network by providing sufficient conditions for uniform exponential stability of a fixed point ~\cite{qin2022vibrational}. 
Furthermore, the work in~\cite{menara2022functional} focuses on stabilizing a nonlinear network of oscillators by providing the patterns of the pairwise relationships between the oscillators’ phases. 
In addition, the work in~\cite{qin2023vibrational} provides sufficient conditions for vibrational inputs to stabilize cluster synchronization of a nonlinear network of oscillators and offers a tractable approach for designing vibrational control.
Finally, for a bistable dynamical system, the authors provide conditions on external perturbations to ensure input-to-state stability~\cite{qin2024analytical}. 
The main limitations of considering a network of nonlinear oscillators or bistable dynamical systems is the inability to estimate the parameters of the model from data and the issue of scalability.  

Fractional-order dynamical networks can be efficiently estimated from data~\cite{gupta2018unknown,gupta2019learning,chatterjee2022learning} and do not suffer from issues of scalability. 
In our prior work, we derived conditions for global asymptotic stability of time-invariant fractional-order systems in discrete-time~\cite{reed2022quantification,reed2023mitigating}. 
Furthermore, we derived two stabilizing feedback strategies by altering the parameters of fractional-order systems and provided the polynomial-time computational complexity to compute these feedback strategies~\cite{reed2023mitigating}.

In this work, we demonstrate the capability of our state-of-the-art stabilizing state feedback control scheme to effectively suppress epileptic activity in 34 out of 35 patients.  
Future work will focus on comparing our state feedback control method with traditional stabilizing state feedback methods for linear time-invariant dynamics.

Patient HUP78, who developed epilepsy from traumatic injury, failed to achieve stabilization in all five of their seizures, including three of their seizures that were initially stable. 
This complete failure suggests that traumatic brain injury-induced structural damage, gliosis, and altered network connectivity~\cite{pitkanen2014epilepsy} may require control strategies different from other epilepsy etiologies.
Additionally, the eight seizures that failed to stabilize exhibited severely ill-conditioned optimization problems, with the median condition numbers~50x larger than those of successful cases (1383 vs 28), which suggests that a feasible solution to the optimization problem presented in~\eqref{eq:optimization_3} may not exist. 
Physiologically, this may indicate that certain seizure dynamics involve such profound network reorganization or high-dimensional complexity that linear coupling modifications may be unable to restore stability~\cite{Scheid2021}.
Importantly, most control failures occurred in seizures with good model fits, indicating that accurate system identification does not guarantee a feasible solution to the stabilizing control problem formulated as an optimization problem in~\eqref{eq:optimization_3}. 
Future work will focus on developing feasibility conditions for solving the stabilizing control problem.  

\subsection*{Limitations}

A potential limitation of this study is the sample size (10 patients, 35 seizures), which may limit statistical power despite large effect sizes. 
Larger cohorts would enable the identification of patient subgroups with distinct seizure dynamics profiles and improve statistical power for detecting group-level patterns.

We observed that the goodness of fit of our model varied across segments, with post-ictal periods being particularly challenging. In our work, we considered a 3 second sliding window to estimate fractional-order dynamical networks. 
In future work, we will investigate the effects of window size selection on estimation. 
A systematic investigation of window size and stride parameters could optimize the temporal resolution for capturing seizure transitions.
Additionally, more work is needed to assess the sensitivity of our estimation scheme to noise and extreme perturbations.  

Our current control method may face real-world implementation challenges. 
In future work, we plan to address the issue of robustness and develop a scheme to overcome delays. 
Furthermore, our current strategy ignores stimulation artifacts and safety constraints, which we plan to address in a future study. 

In this study, we focused on analyzing eigenvalues and fractional-order exponents ($\alpha$) of fractional-order dynamical networks, which show significant promise. 
In future work, we plan to analyze eigenvectors, which may help to identify the spatial patterns of network reorganization across the seizure states and reveal which electrode drives transitions between states, which has the potential to improve targeting for neurostimulation. 
  

Future work will focus on developing conditions to guarantee a feasible control solution by verifying stabilizability with state-feedback. Certain network configurations that are well-characterized by fractional-order models may be fundamentally resistant to state-feedback interventions due to structural constraints, pathological states, or intrinsic nonlinearities that emerge during control attempts~\cite{Scheid2021}. 
The numerical ill-conditioning observed in failed cases may thus reflect genuine physiological properties of these seizures rather than modeling inadequacies.
The heterogeneity in control outcomes also points toward the need for patient-specific or etiology-specific controller designs. 
Patients with structural lesions, certain epilepsy etiologies, or particular network architectures may require alternative control strategies, such as nonlinear control, higher-order perturbations, or targeting different network nodes. 
Larger studies grouping patients by clinical characteristics may reveal which characteristics are more favorable to state-feedback control versus those requiring more sophisticated interventions~\cite{Stacey2008, Scheid2021}.
Finally, we evaluated our control framework \textit{in silico} simulations using estimated models from real-world iEEG data. 
We plan to test our control framework \textit{in vivo} and \textit{in vitro} using animal models and eventually in a clinical setting, which would help us assess the real-world feasibility and efficacy of fractional-order model-based neurostimulation for seizure suppression.

\section*{Methods}

\subsection*{Epileptic Intracranial EEG HUP and Mayo Dataset}

We analyzed intracranial EEG (iEEG) recordings from 10 patients with medically refractory epilepsy, which are available through the International Epilepsy Electrophysiology Portal~\cite{wagenaar2013multimodal}. 
These patients underwent subdural electrode implantation for presurgical evaluation after a noninvasive assessment suggested focal, surgically amenable epilepsy. Subdural grid and strip electrodes (2.3 mm diameter, 10 mm spacing) were placed based on clinical indication~\cite{ashourvan2020model}. Table \ref{tab:patients} summarizes the clinical characteristics of these patients with all abbreviations defined in the table caption.

There were a total of 35 seizure blocks across the 10 patients. Nine patients experienced complex partial (CP) or complex partial with secondary generalized tonic-clonic (CP+GTC) seizures, and one patient exhibited simple partial (SP) seizures. All seizures had a focal, localized onset, with the time of onset annotated by expert clinicians~\cite{khambhatiDynamicNetworkDrivers2015,litt2001epileptic}. Six patients were recorded at the Hospital of the University of Pennsylvania (HUP) and four at the Mayo Clinic (MAYO), with sampling rates of 512\,Hz and 500\,Hz, respectively. Detailed electrode configurations and recording specifications are described in~\cite{ashourvan2020model}.
\begin{table}[h!]
\footnotesize
\centering
\caption{\footnotesize
Clinical characteristics of the 10 epilepsy patients in our dataset. Onset and surgery refer to age at first seizure and age at phase II monitoring, respectively. Etiology is clinically determined. Seizure types: CP = complex partial, GTC = generalized tonic-clonic, and SP = simple partial. The seizures column indicates how many seizures are analyzed per patient. Imaging: L = lesional, NL = non-lesional. Outcome: ENGEL classification (I-IV) or ILAE classification where indicated, NR = not reported, NF = no follow-up. All seizures are localized with defined seizure onset zones.
}
\begin{tabularx}{\textwidth}{l c c X X X c c c}
\toprule
Patient & Sex & Age (Years) & Seizure Onset & Etiology & Seizure Type & Seizures & Imaging & Outcome \\
 & & (Onset/Surgery) & & & & & & \\
\midrule
HUP64 & M & 0.3/20 & Left frontal & Dysplasia & CP+GTC & 1 & L & ENGEL-I \\
HUP68 & F & 15/26 & Right temporal & Meningitis & CP, CP+GTC & 5 & NL & ENGEL-I \\
HUP70 & M & 10/32 & Left perirolandic & Cryptogenic & SP & 8 & L & NR \\
HUP72 & F & 11/27 & Bilateral left & Mesial temporal sclerosis & CP+GTC & 1 & L & NR \\
HUP78 & M & 00/54 & Anterior left temporal & Traumatic Injury & CP & 5 & L & ENGEL-III \\
HUP86 & F & 18/25 & Left temporal & Cryptogenic & CP+GTC & 2 & NL & ENGEL-II \\
MAYO010 & F & 00/13 & Left frontal & Neonatal injury & CP+GTC & 2 & L & NF \\
MAYO011 & F & 10/34 & Right Mesial frontal & Unknown & CP & 2 & NL & NF \\
MAYO016 & F & 05/36 & Right temporal orbitofrontal & Unknown & CP+GTC & 3 & NL & ILAE-IV \\
MAYO020 & F & 05/10 & Right frontal & Unknown & CP+GTC & 4 & NL & ILAE-IV \\
\bottomrule
\end{tabularx}
\label{tab:patients}
\end{table}

 \subsection*{Preprocessing}
The iEEG data were pre-processed according to the details described in~\cite{ashourvan2020model}, which included 0.1 Hz high-pass filtering, removal of artifact electrodes annotated by clinical experts, and epileptic brain state segmentation. 
The first 20 seconds of each seizure recording were clinically annotated as pre-ictal, and the last 20 seconds of each seizure recording were clinically annotated as post-ictal. 
The interval between pre-ictal and post-ictal of each seizure recording was clinically annotated as ictal. 
Finally, the snapshot of interictal data was randomly selected at least three hours before or after any seizure event and divided into 100-second snapshots, with each patient having at least as many interictal snapshots as ictal snapshots. 
From interictal, pre-ictal, ictal, and post-ictal snapshots, we analyzed 20 seconds of data to ensure a consistent comparison across epileptic brain states.

For our analysis, we performed additional processing to prepare the data for fractional-order dynamical network modeling. 
Each signal from every electrode was mean-centered by subtracting its average during 3-second windows. 
We chose a 3-second window length to balance temporal resolution with sufficient data for model fitting. 

To each 20-second epileptic brain state segment, we applied a 3-second sliding window with a 1-second stride to track temporal changes across time. 
Each 20-second segment yielded 18 windows (windows starting at 0, 1, ..., 17 seconds), with the final window spanning 17-20 seconds. 
These windowed data segments served as inputs to our FOS modeling pipeline, which we used to compute stability and multi-scale properties across time.

\subsection*{Fractional-Order Dynamical Network Modeling Framework}
The evolution of multivariate brain dynamics can be expressed in the following form:
\begin{equation}\label{eq:fractional_sys}
 \Delta^{\alpha}{x}[k] = A {x}[k-1],
\end{equation}
where ${x}[k] = (x_1[k], x_2[k], \ldots, x_N[k])^{{T}}$ is a state vector of the $N$ pre-processed iEEG recordings at discrete time step $k \in \mathbb{N}$. 
$A \in \mathbb{R}^{N \times N}$ is a matrix that models the spatial coupling between states. 
The multi-scale behavior is described using a discrete-time fractional derivative operator $\Delta^{\alpha}$, which is the \mbox{Gr\"unwald-Letnikov} discretization of the fractional derivative~(Chpt.~2,~\cite{podlubny1998fractional}). The vector of fractional-order exponents ${\alpha} = (\alpha_1, \alpha_2, \ldots, \alpha_N)^{\mathsf{T}}$ computes the multi-scale property for each recording, which determines how strongly each measurement depends on its past measurements. 
Smaller values of $\alpha$ indicate a stronger long-range temporal dependency. 
A value of $\alpha = 1$ corresponds to a standard first-order linear time-invariant dynamical system with no multi-scale behavior.
This relationship is described for each electrode $i$ by:
\begin{equation}\label{eq:single_state_DTFOS}
    \Delta^{\alpha_{i}} x_{i}[k] = \sum\limits_{j=0}^{k} \Psi(\alpha_{i}, j) x_{i}[k-j],
\end{equation}
where $\Psi(\alpha_{i}, j)$ are the Grünwald-Letnikov weights defined as:
\begin{equation}\label{eq:GL_weights}
    \Psi(\alpha_{i}, j) = \begin{cases}
        1, & j = 0 \\
        \frac{\Gamma(j - \alpha_i)}{\Gamma(-\alpha_i) \Gamma(j + 1)}, & j \geq 1
    \end{cases}
\end{equation}
These weights determine the contribution of each past state, with the influence decaying according to the fractional order $\alpha_i$.

\subsection*{Model Estimation}
The fractional-order exponents $\alpha$ were estimated using a Haar wavelet transform. 
For each electrode, we computed the log-variance of the Haar wavelet 
coefficients at $J = \lfloor \log_2(K) \rfloor$ scales, where $K$ is the window length, and performed a linear regression between the log-scale and log-variance.
The slope of this regression is divided by two and gives us the estimated $\alpha_i$ values~\cite{flandrin1992wavelet, abry1998wavelet}. 
These $\alpha_i$ values are then used to compute the Grünwald-Letnikov approximation of the discrete-time fractional derivative $\Delta^{\alpha_i} x_i[k]$ using Eq.~\eqref{eq:single_state_DTFOS}, with a finite memory of 20 past samples. 

The system matrix A was initially estimated using ordinary least squares (OLS) by solving $A = Z X^{\mathsf{T}} (X X^{\mathsf{T}})^{-1}$, where $X \in \mathbb{R}^{N \times K}$ contains the state vectors with the $k$-th column being $x[k-1]$ for $k = 1, 2, \ldots, K$, with K denoting the number of time windows, and $Z \in \mathbb{R}^{N \times K}$ contains the corresponding fractional derivatives with the $k$-th column being $\Delta^{\alpha} x[k]$.

To improve the estimation of $A$ and account for potential unmeasured influences, we used an iterative estimation approach~\cite{gupta2018unknown} to estimate the parameters and unknown inputs. 
We augmented Eq.~\eqref{eq:fractional_sys} to include an input term:
\begin{equation}
    \Delta^{\alpha} x[k] = A x[k-1] + B u[k],
\end{equation}
where $u[k]$ represents any external inputs to the system from unmeasured sources (e.g., brain regions outside the recording area, subcortical inputs, or modeling errors). 
The matrix $B$ was initialized heuristically from $A^{(0)}$, which is the estimated $A$ matrix during the first iteration, by selecting columns that represent the most influential electrodes in the network, assuming half the recorded electrodes could receive external inputs. 
For each time iteration $i$, we estimated the inputs by solving a least absolute shrinkage and selection operator (LASSO) problem:
\begin{equation}
    \min_{u[k]} \frac{1}{2} \| z[k] - A^{(i)} x[k-1] - B u[k] \|_2^2 + \mu \| u[k] \|_1,
\end{equation}
using the Alternating Direction Method of Multipliers (ADMM)~\cite{Boyd2011ADMM}, with regularization parameter $\mu = 0.5$. After identifying these contributions, we re-estimated the system matrix:
\begin{equation}
    A^{(i+1)} = (Z - BU) X^{\mathsf{T}} (X X^{\mathsf{T}})^{-1},
\end{equation}
where $U$ contains all estimated inputs. In the final model, only one iteration was used as additional iterations did not significantly improve the mean squared error. 

To analyze the network coupling structure with adjusted diagonal terms, we computed $A_0 = A - D$, where $D = \text{diag}(v)$ with diagonal elements $v_i = \frac{\Gamma(1-\alpha_i)}{\Gamma(-\alpha_i)}$ for $\alpha_i \neq 0$ and $v_i = 1$ for $\alpha_i = 0$.
The matrix $A_0$ represents the effective inter-channel coupling after accounting for the diagonal contribution from fractional differentiation. We computed all the eigenvalues of $A_0$ for each window to characterize the network dynamics. These eigenvalues, along with the fractional order $\alpha$ values, served as parameters to differentiate between seizure states. With the refined $A^{(1)}$ and $\alpha$, we can now reconstruct the signals as described below.

\subsection*{Artificial Signal Simulation}
The model quality was evaluated by simulating signals using the estimated parameters $A$ and $\alpha$. 
Starting from Eq.~\eqref{eq:fractional_sys} and expanding the fractional derivative using the Grünwald-Letnikov weights from Eq.~\eqref{eq:GL_weights} with $\Psi(\alpha_i, 0) = 1$:
\begin{equation}
\Delta^{\alpha_i}x_i[k] = \sum_{j=0}^{k}\Psi(\alpha_i,j)x_i[k-j] = x_i[k] + \sum_{j=1}^{k}\Psi(\alpha_i,j)x_i[k-j] 
\end{equation}

Recall that for a scalar entry of the state 

\begin{equation}
    \Delta^{\alpha_i}x_i[k] = (Ax[k-1])_i
\end{equation}

Rearranging to solve for $x[k]$, we obtain:
\begin{equation}
x_i[k] = A x[k-1]_i - \sum_{j=1}^{\min(k,M)} \Psi(\alpha_i,j) x_i[k-j]
\label{eq:reconstruction}
\end{equation}
where $(A x[k-1])_i$ represents the multi-channel coupling and the summation captures the fractional multi-scale contribution, limited to $M = 5$ past samples. Starting from initial conditions $x[0]$, we applied Eq.~\eqref{eq:reconstruction} recursively to generate multi-step-ahead predictions across all electrodes. 

We quantified reconstruction quality by computing the coefficient of determination ($R^2$) between reconstructed and preprocessed iEEG signals for each electrode in each time window. 
For statistical comparisons between seizure states, we excluded snapshots where fewer than 60\% of electrodes achieved a mean $R^2 \geq 0.5$ to ensure at least half the variance was captured in the original signal. However, all snapshots were retained for visualizing the fractional-order exponents ($\alpha$) and eigenvalue distributions.

\subsection*{Analysis of Multi-scale and Stability Properties}

For statistical analysis, we examined fractional-order exponents ($\alpha$) and eigenvalues across epileptic brain states through population pooling and within-patient analysis using the Kolmogorov-Smirnov (KS) test and Cohen's $d$ statistic. Only segments that passed the model fit criteria were included, as described in the previous section. 

The KS test measures the maximum difference between cumulative distribution functions, through a $p$-value and a KS statistic, which ranges from 0 (identical distributions) to 1 (completely separated)~\cite{wilcox2011introduction}. Cohen's $d$ quantifies effect size as the standardized mean difference in units of pooled standard deviation, where $|d| \geq 0.2$, $\geq 0.5$, and $\geq 0.8$ represent small, medium, and large effects, respectively~\cite{wilcox2011introduction}. 

For the population-pooled approach, we aggregated all values within each brain state and visualized the distributions using violin plots. We computed pairwise KS tests and Cohen's $d$ value for all segment comparisons, which are shown in a heatmap, while the $p$-values are shown in the violin plots. Large sample sizes often guarantee statistical significance from $p$-values; therefore, effect size metrics (KS statistic and Cohen's $d$) provide a more informative approach for assessing differences between brain states.

For the within-patient approach, we generated violin plots for each patient by aggregating all available segments. We then computed KS statistics and Cohen's $d$ for each pairwise segment comparison within individual patients. These values were averaged within each patient and then averaged across all patients to produce a summary heatmap. The $p$-values were not reported for this analysis as the averaged effect sizes do not correspond to a single hypothesis test. This approach quantifies whether effects are consistent across individuals while accounting for inter-patient variability.

\subsection*{Stabilizing Controller Design}
A stabilizing controller was designed to stabilize the fractional-order dynamical network model during ictal periods, based on the stability framework for fractional-order systems ~\cite{reed2023mitigating}. In general, stability describes the long-term behavior of signals. A system is said to be \emph{stable} if all of the signals decay to zero as time goes to infinity~\cite{hespanha2018linear}. Hence, we defined stability for fractional-order systems. 
\begin{definition}\label{def:stability}
\textit{Stability of Fractional-Order Systems: (Theorem~1, \cite{reed2023mitigating})} A fractional-order system~\eqref{eq:fractional_sys} is said to be stable \textit{if and only if} for $A_0:=A-D(\alpha,1)$, where
$D(\alpha,j) = \left[\begin{smallmatrix} \psi(\alpha_{1},j) & 0 & \dots & 0 \\
0 & \psi(\alpha_{2},j) & \dots & 0 \\ 
0 & \vdots & \ddots & 0 \\
0 & 0 & \dots & \psi(\alpha_{n},j)
\end{smallmatrix}\right],$ we have $|\lambda|<1$ for all $\lambda\in\sigma(A_0)$, where $\sigma(A_0)$ is the set of eigenvalues of matrix $A_0$. 
\end{definition}

Given $(A,\alpha)$, we seek a coupling matrix $\tilde{A}$ that satisfies the following:
\begin{equation}\label{eq:optimization_1}
\begin{split}
    \min_{\tilde{A}\in\mathbb{R}^{N \times N}}\quad & \|\tilde{A}\|_{0} \\
     \text{s.t. } & (A + \tilde{A}, \alpha) \text{ is globally} \\  
     &\text{asymptotically stable},
\end{split}
\end{equation}
where $\|\cdot\|_0$ represents the zero quasi-norm, which counts the number of non-zero entries in a matrix. This seeks the sparsest possible control intervention, modifying the fewest connections in the brain network. However, finding the globally optimal solution to this problem requires exhaustively searching through all possible combinations of non-zero entries, which becomes computationally intensive even for moderately sized networks.

Since this problem is computationally intractable, we solve the convex relaxation using the $\ell_1$ norm:
\begin{equation}\label{eq:optimization_2}
\begin{split}
    \min_{\tilde{A}\in\mathbb{R}^{N \times N}}\quad & \|\tilde{A}\|_{1} \\
     \text{s.t. } & \rho( A+\tilde{A}-D(\alpha,1))<1,
\end{split}
\end{equation}
where $\|\cdot\|_1 = \sum_{i,j} |[\tilde{A}]_{ij}|$ is the element-wise $\ell_1$ norm, which serves as the tightest convex relaxation of the $\ell_0$ norm and promotes sparsity while remaining computationally tractable, and $\rho(\cdot)$ denotes the spectral radius (largest absolute eigenvalue) of a matrix.

The solution to equation~\eqref{eq:optimization_2} is given by $\tilde{A}=LP^{-1}$, where $P$ and $L$ are found by solving the following convex optimization problem:
\begin{equation}\label{eq:optimization_3}
\begin{split}
   \min_{P,L} \quad & \|P\|_{1}+\|L\|_{1} \\
     \text{s.t. } & \begin{bmatrix}
     P & PA_0^T + L^T \\
     A_0P+L & P
     \end{bmatrix} \succ 0, \\
     & P \succ I
\end{split}
\end{equation}

This controller was applied to ictal windows, starting at seizure onset.
We evaluated controller effectiveness by comparing the eigenvalues of $A_0$ matrix (uncontrolled) and $A_0 + \tilde{A}$ matrix (controlled) during the ictal period. Additionally, to visualize the effect of control on signal dynamics, we reconstructed signals using Eq.~\eqref{eq:reconstruction}, substituting $A_{\text{ctrl}} = A + \tilde{A}$ for $A$, to visualize the effect of control on signal amplitudes.

\section*{Acknowledgments}

Code is available at https://github.com/Yaoyuewang/fractional-control-epilepsy. Data are available from the International Epilepsy Electrophysiology Portal (IEEG Portal).

\newpage

\newpage
\section*{Supplementary Material}
Table~\ref{tab:excluded_segments} shows the excluded segments and the corresponding average $R^2$ across all channels. Negative $R^2$ values indicate the model performance was worse than a mean predictor. All 8 ictal segments from patient HUP70 passed the $R^2$ criterion, but their duration was less than 20 seconds, so we excluded them from the pairwise comparison. 

\begin{table}[h!]
\centering
\caption{Excluded segments summary with mean R² values and percentage of channels meeting inclusion criteria}
\label{tab:excluded_segments}
\begin{tabular}{lcccc}
\hline
Patient & Segment Type & Count & Mean R² & \% Good Channels \\
\hline
HUP64   & Post-ictal    & 1 & 0.345  & 17.4\% \\
HUP68   & Post-ictal    & 3 & 0.313  & 8.0\%  \\
HUP78   & Pre-ictal     & 1 & 0.429  & 33.7\% \\
        & Post-ictal    & 1 & 0.433  & 32.6\% \\
        & Inter-ictal   & 5 & $-$0.192 & 13.5\% \\
HUP86   & Post-ictal    & 2 & $-$0.412 & 0.0\%  \\
MAYO016 & Post-ictal    & 1 & 0.328  & 24.1\% \\
\hline
\end{tabular}
\end{table}

\begin{table}[h!]
\centering
\caption{Patient-specific fractional order exponents ($\alpha$) for each epileptic brain state. Values represent median (IQR) across all seizures for each patient-segment combination. Only segments with good model fit ($R^2 > 0.5$ in $\geq60\%$ of channels) are included. N/A: segment excluded due to poor model fit.}
\label{tab:alpha_patient_good_fit}
\begin{tabular}{lcccc}
\hline
\textbf{Patient} & \textbf{Inter-ictal} & \textbf{Pre-ictal} & \textbf{Ictal} & \textbf{Post-ictal} \\
\hline
HUP64    & 0.75 (0.68--0.81) & 0.56 (0.51--0.61) & 0.60 (0.55--0.65) & N/A \\
HUP68    & 0.80 (0.71--0.89) & 0.70 (0.58--0.84) & 0.63 (0.54--0.72) & 0.67 (0.55--0.80) \\
HUP70    & 0.75 (0.66--0.84) & 0.76 (0.66--0.87) & N/A & 0.75 (0.66--0.88) \\
HUP72    & 0.62 (0.55--0.72) & 0.99 (0.93--1.04) & 0.97 (0.91--1.05) & 0.79 (0.73--0.85) \\
HUP78    & N/A & 0.51 (0.40--0.62) & 0.46 (0.36--0.56) & 0.50 (0.40--0.62) \\
HUP86    & 0.75 (0.68--0.82) & 0.39 (0.32--0.48) & 0.38 (0.32--0.45) & N/A \\
MAYO010  & 0.73 (0.64--0.80) & 0.84 (0.74--0.94) & 0.75 (0.63--0.85) & 0.81 (0.73--0.88) \\
MAYO011  & 0.68 (0.59--0.77) & 0.67 (0.57--0.77) & 0.68 (0.57--0.80) & 1.02 (0.90--1.13) \\
MAYO016  & 0.78 (0.69--0.87) & 0.61 (0.51--0.72) & 0.66 (0.54--0.78) & 0.99 (0.83--1.16) \\
MAYO020  & 0.72 (0.62--0.82) & 0.80 (0.69--0.90) & 0.80 (0.70--0.91) & 0.91 (0.79--1.03) \\
\hline
\end{tabular}
\end{table}

\begin{table}[h!]
\centering
\caption{Patient-specific eigenvalues for each epileptic brain state. Values represent median (IQR) across all seizures for each patient-segment combination. Only segments with good model fit ($R^2 \geq 0.5$ in $\geq60\%$ of channels) are included. N/A: segment excluded due to poor model fit.}
\label{tab:eigen_patient_good_fit}
\begin{tabular}{lcccc}
\hline
\textbf{Patient} & \textbf{Inter-ictal} & \textbf{Pre-ictal} & \textbf{Ictal} & \textbf{Post-ictal} \\
\hline
HUP64    & 0.80 (0.77--0.84) & 0.73 (0.71--0.75) & 0.74 (0.72--0.76) & N/A \\
HUP68    & 0.83 (0.78--0.88) & 0.77 (0.74--0.84) & 0.76 (0.73--0.78) & 0.75 (0.67--0.82) \\
HUP70    & 0.80 (0.76--0.85) & 0.80 (0.76--0.87) & N/A               & 0.80 (0.76--0.87) \\
HUP72    & 0.74 (0.72--0.77) & 0.96 (0.90--1.01) & 0.96 (0.90--1.02) & 0.62 (0.54--0.75) \\
HUP78    & N/A               & 0.73 (0.70--0.76) & 0.72 (0.69--0.76) & 0.73 (0.69--0.76) \\
HUP86    & 0.80 (0.76--0.83) & 0.72 (0.70--0.75) & 0.72 (0.70--0.74) & N/A \\
MAYO010  & 0.77 (0.74--0.81) & 0.86 (0.79--0.93) & 0.80 (0.75--0.86) & 0.80 (0.69--0.87) \\
MAYO011  & 0.77 (0.74--0.80) & 0.77 (0.74--0.80) & 0.77 (0.73--0.81) & 0.96 (0.80--1.08) \\
MAYO016  & 0.82 (0.77--0.87) & 0.74 (0.71--0.77) & 0.75 (0.72--0.80) & 0.93 (0.81--1.10) \\
MAYO020  & 0.78 (0.74--0.84) & 0.82 (0.77--0.89) & 0.82 (0.77--0.90) & 0.86 (0.77--0.98) \\
\hline
\end{tabular}
\end{table}

\begin{table}[t!]
\centering
\caption{Patient-specific fractional order exponents ($\alpha$) and eigenvalues for excluded segments with poor model fit ($R^2 < 0.5$ in $\geq40\%$ of channels). Values represent median (IQR) across all seizures for each patient-segment combination.}
\label{tab:bad_fit_summary}
\begin{tabular}{lcccc}
\hline
\textbf{Patient} & \textbf{Segment} & \textbf{$\alpha$} & \textbf{Eigenvalue} \\
\hline
HUP64    & Post-ictal   & 0.27 (0.20--0.38) & 0.52 (0.44--0.60) \\
HUP68    & Post-ictal   & 0.35 (0.26--0.44) & 0.58 (0.48--0.66) \\
HUP78    & Inter-ictal   & 0.11 ($-$0.03--0.27) & 0.84 (0.73--1.01) \\
         & Pre-ictal    & 0.33 (0.24--0.44) & 0.75 (0.69--0.79) \\
         & Post-ictal   & 0.27 (0.18--0.38) & 0.77 (0.70--0.82) \\
HUP86    & Post-ictal   & $-$0.06 ($-$0.24--0.04) & 0.90 (0.79--1.20) \\
MAYO016  & Post-ictal   & 0.68 (0.58--0.78) & 0.73 (0.66--0.81) \\
\hline
\end{tabular}
\end{table}

\section*{Analysis of Poor Model Fits}
For $\alpha$, HUP78 showed markedly lower values during inter-ictal states (median = 0.11, IQR: $-$0.03--0.27) compared to the inter-ictal median of 0.73 observed across good fits. HUP86 exhibited negative $\alpha$ values during post-ictal states (median = $-$0.06, IQR: $-$0.24--0.04). Most excluded post-ictal segments demonstrated substantially reduced $\alpha$ values, except for MAYO016 (median = 0.68).

Excluded eigenvalue segments showed greater heterogeneity across patients. HUP78's inter-ictal segment displayed considerably wider variability (IQR: 0.73--1.01) compared to inter-ictal segments with good fits. Post-ictal eigenvalues were notably lower in HUP64 (median = 0.52) and HUP68 (median = 0.58), both falling outside the typical range observed in good-fit post-ictal segments. These extreme values likely contributed to the poor model fits, suggesting fundamentally different underlying dynamics in these excluded segments.

\section*{Patient-Specific Parameter Distributions}
Figures~\ref{fig:alpha_summary_good_indiv} and \ref{fig:eigen_summary_good_indiv} show individual patient distributions for good fits, while Figures~\ref{fig:alpha_summary_indiv} and \ref{fig:eigen_summary_indiv} include all data.

\begin{figure}[t!]
    \centering
    \includegraphics[width=\textwidth]{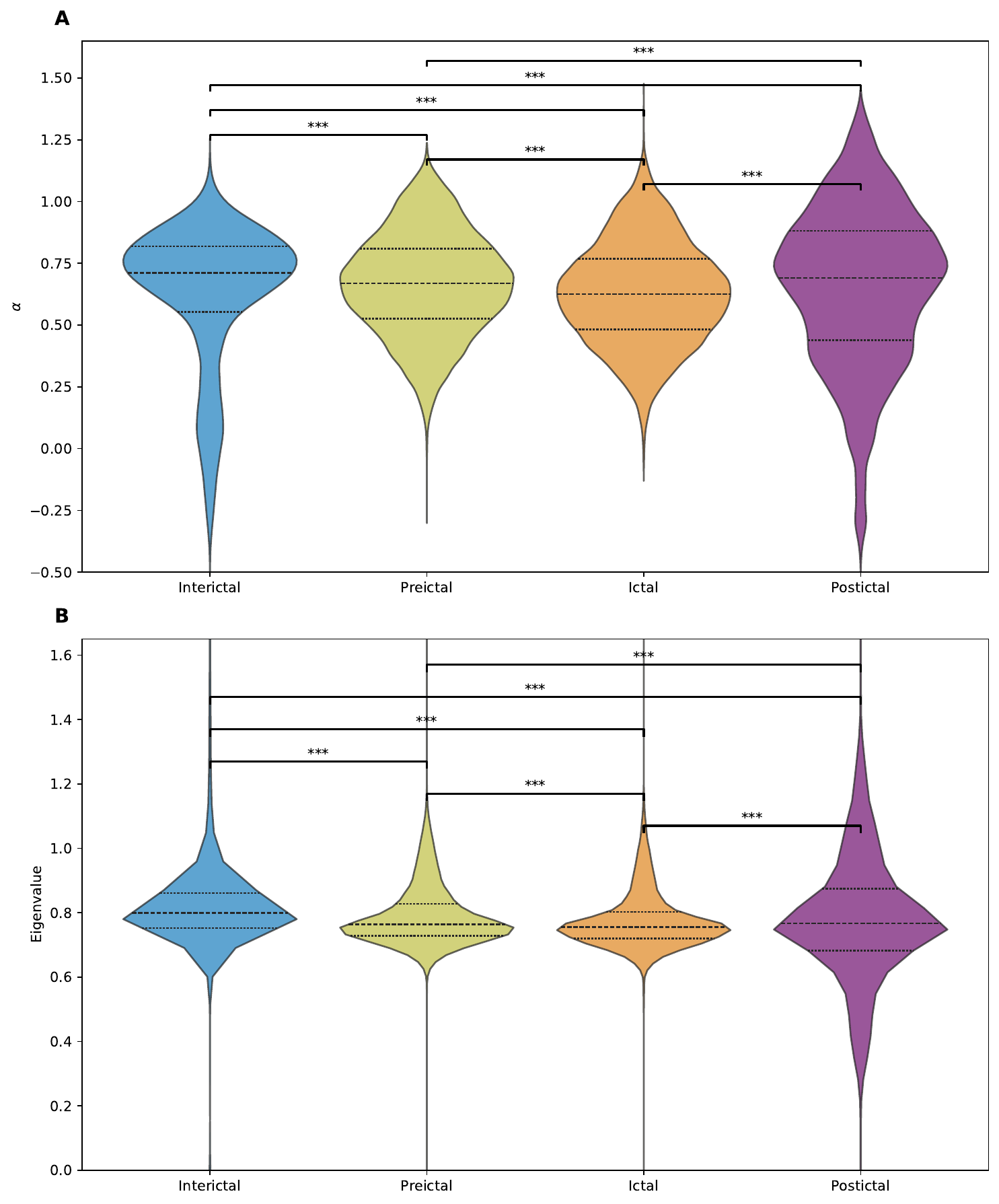}
    \caption{Violin plots of fractional-order system parameters across seizure segments (inter-ictal, pre-ictal, ictal, post-ictal). (A) Fractional-order exponents $\alpha$ characterizing long-term memory of iEEG signals. (B) Eigenvalues from $A_0$ matrix showcasing network dynamics. Data are pooled from all segments, including those with poor model fits.}
    \label{fig:alpha_eigen_summary_all}
\end{figure}

\begin{figure}[t!]
    \centering
    \includegraphics[width=\textwidth]{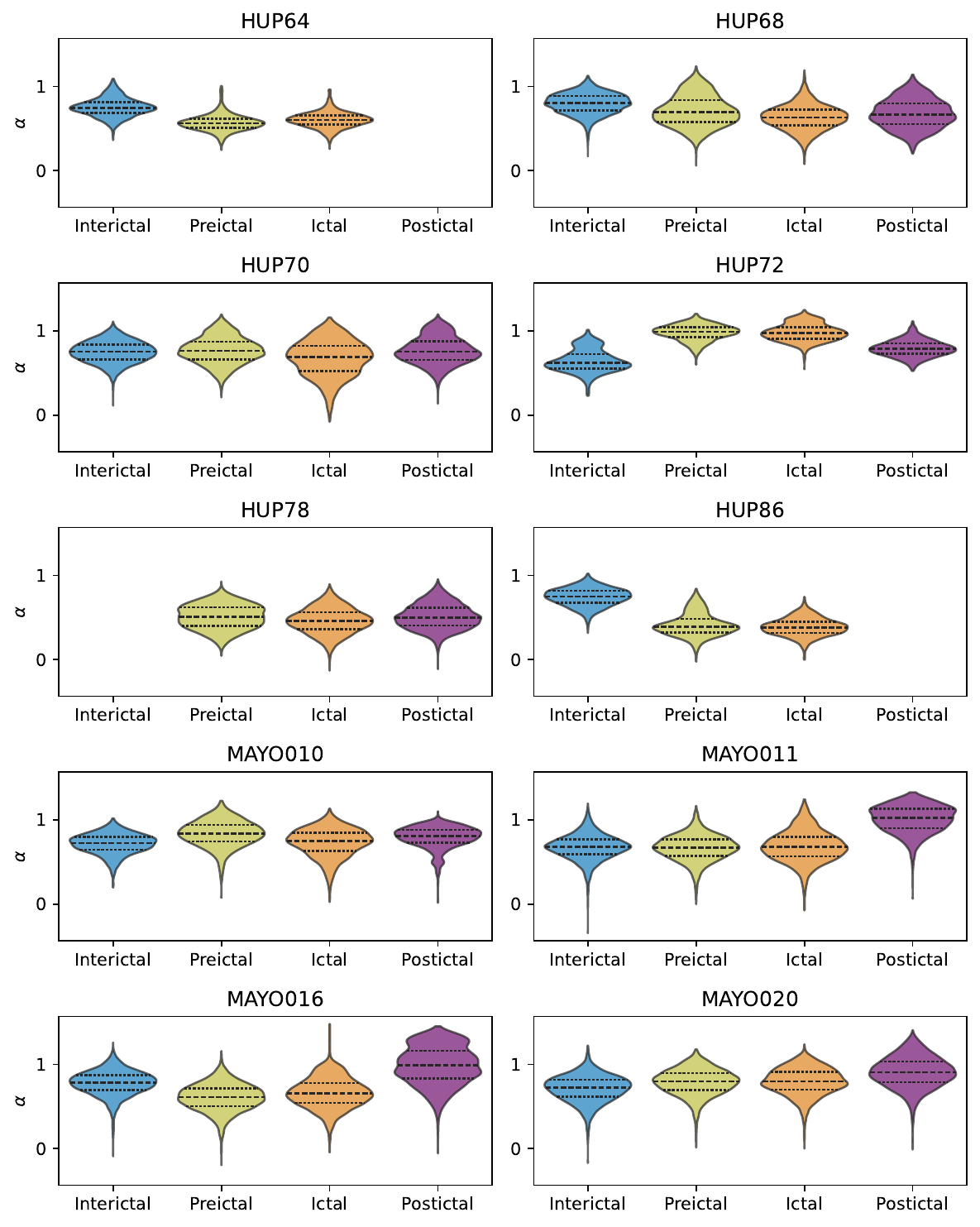}
    \caption{Violin plots of the distribution of fractional-order exponents ($\alpha$) across segments for each patient from only those with good model fit ($R^2 \geq 0.5$ in $\geq60\%$ of channels).}
    \label{fig:alpha_summary_good_indiv}
\end{figure}

\begin{figure}[t!]
    \centering
    \includegraphics[width=\textwidth]{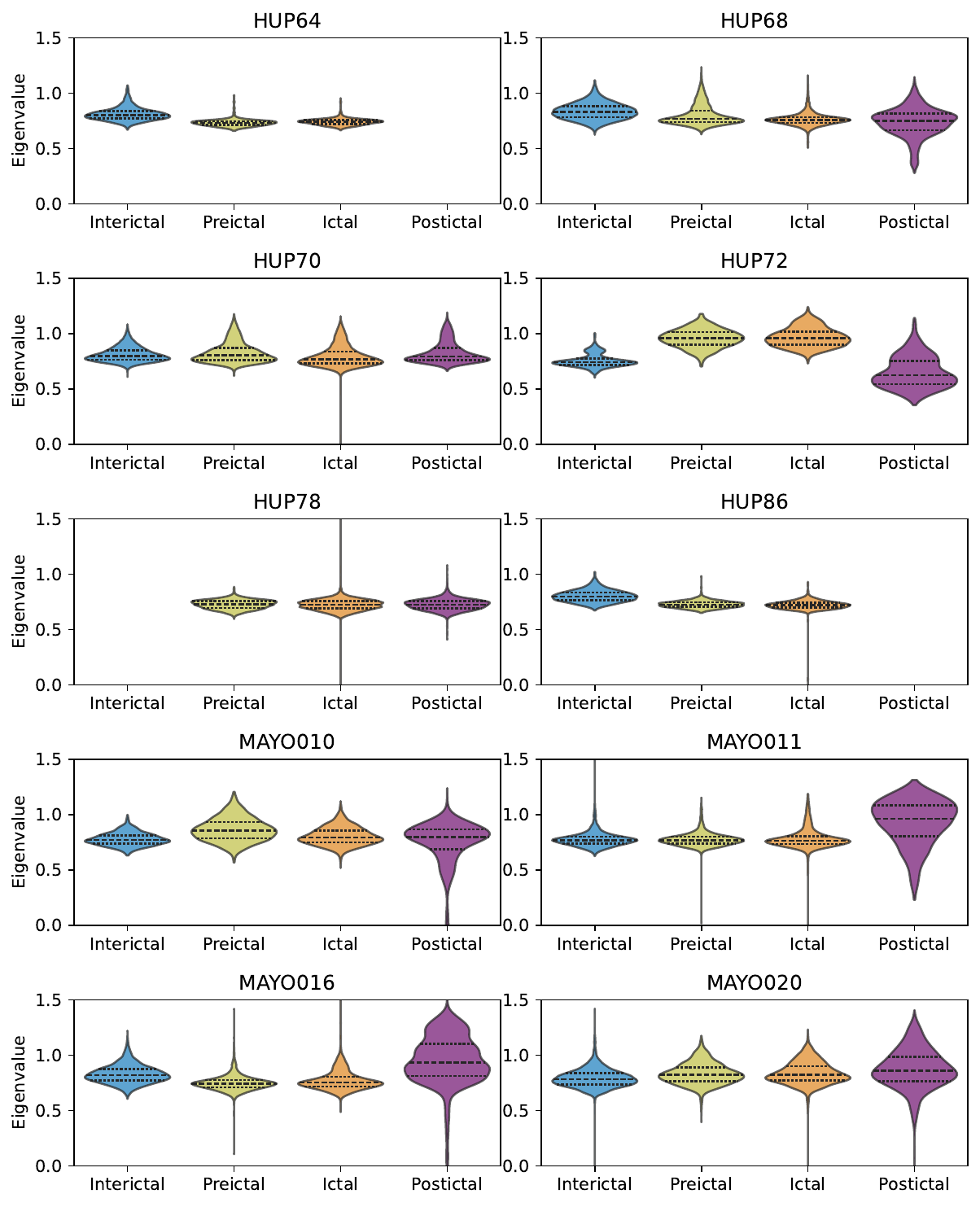}
    \caption{Violin plots of the distribution of eigenvalues across  segments for each patient from only those with good model fit ($R^2 \geq 0.5$ in $\geq60\%$ of channels)}
    \label{fig:eigen_summary_good_indiv}
\end{figure}

\begin{figure}[t!]
    \centering
    \includegraphics[width=\textwidth]{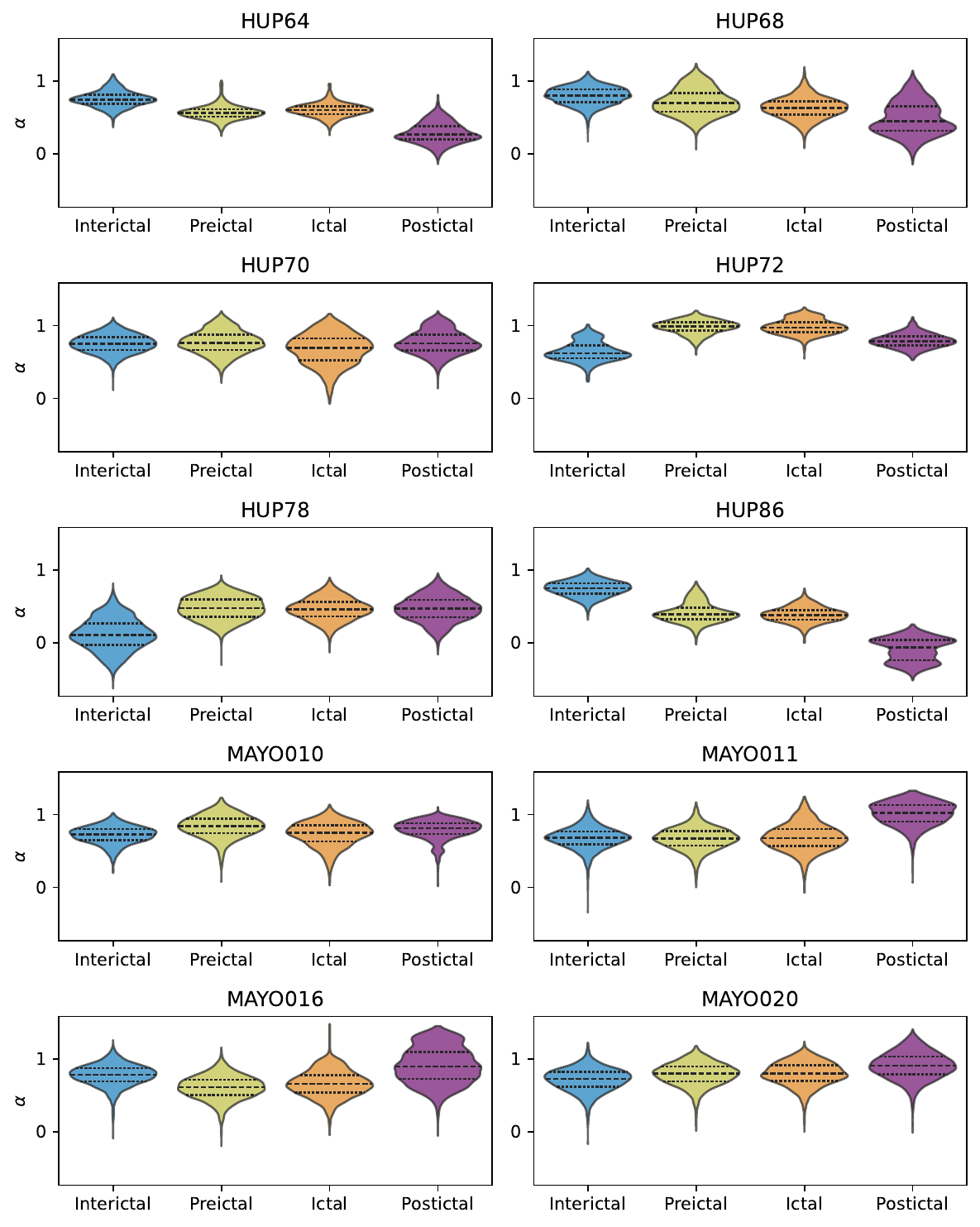}
    \caption{Violin plots of the distribution of fractional-order exponents ($\alpha$) across segments for each patient, including all data regardless of model fit quality.}
    \label{fig:alpha_summary_indiv}
\end{figure}

\begin{figure}[t!]
    \centering
    \includegraphics[width=\textwidth]{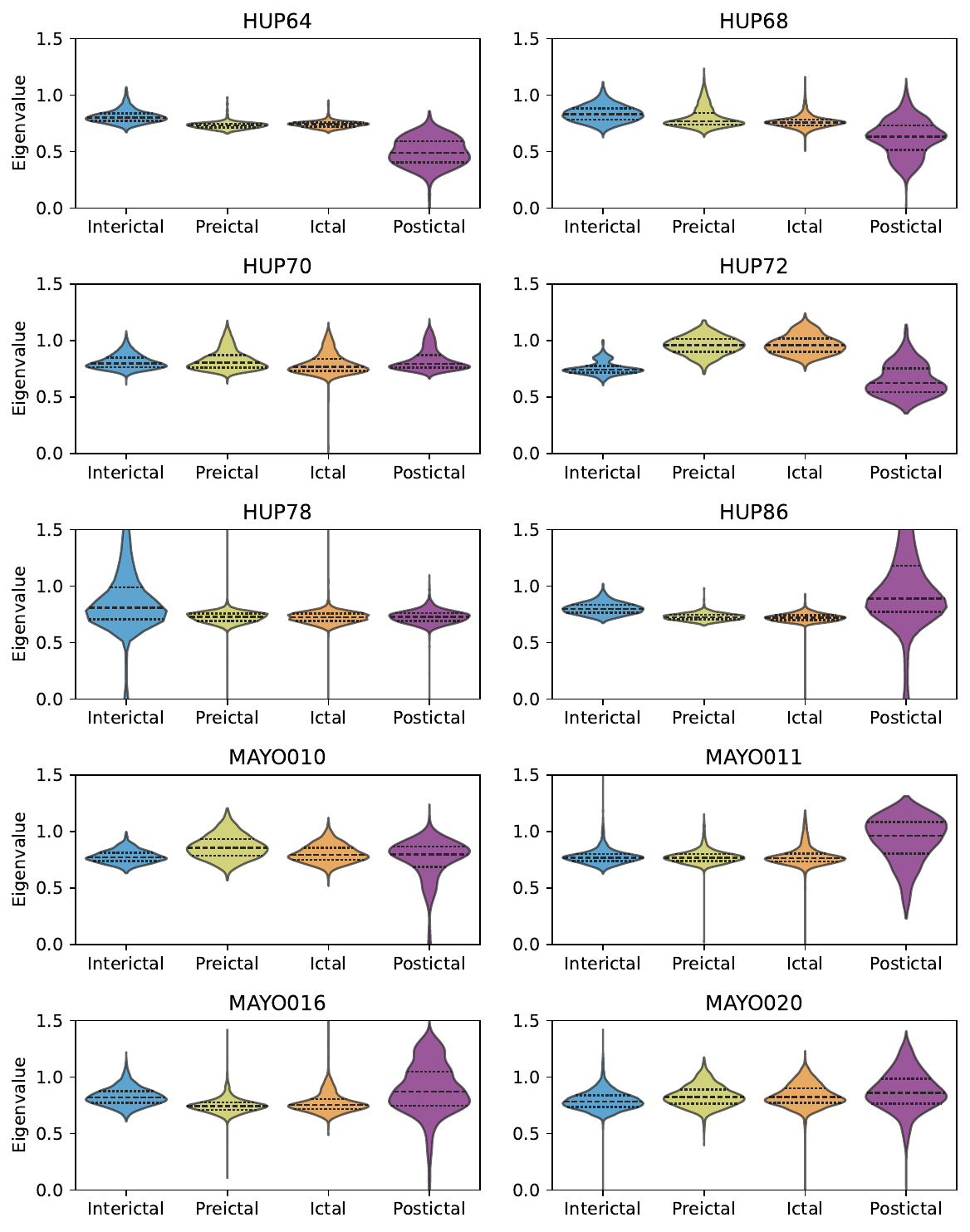}
    \caption{Violin plots of the distribution of eigenvalues across segments for each patient, including all data regardless of model fit quality.}
    \label{fig:eigen_summary_indiv}
\end{figure}

\begin{figure}[t!]
    \centering
    \includegraphics[width=\textwidth]{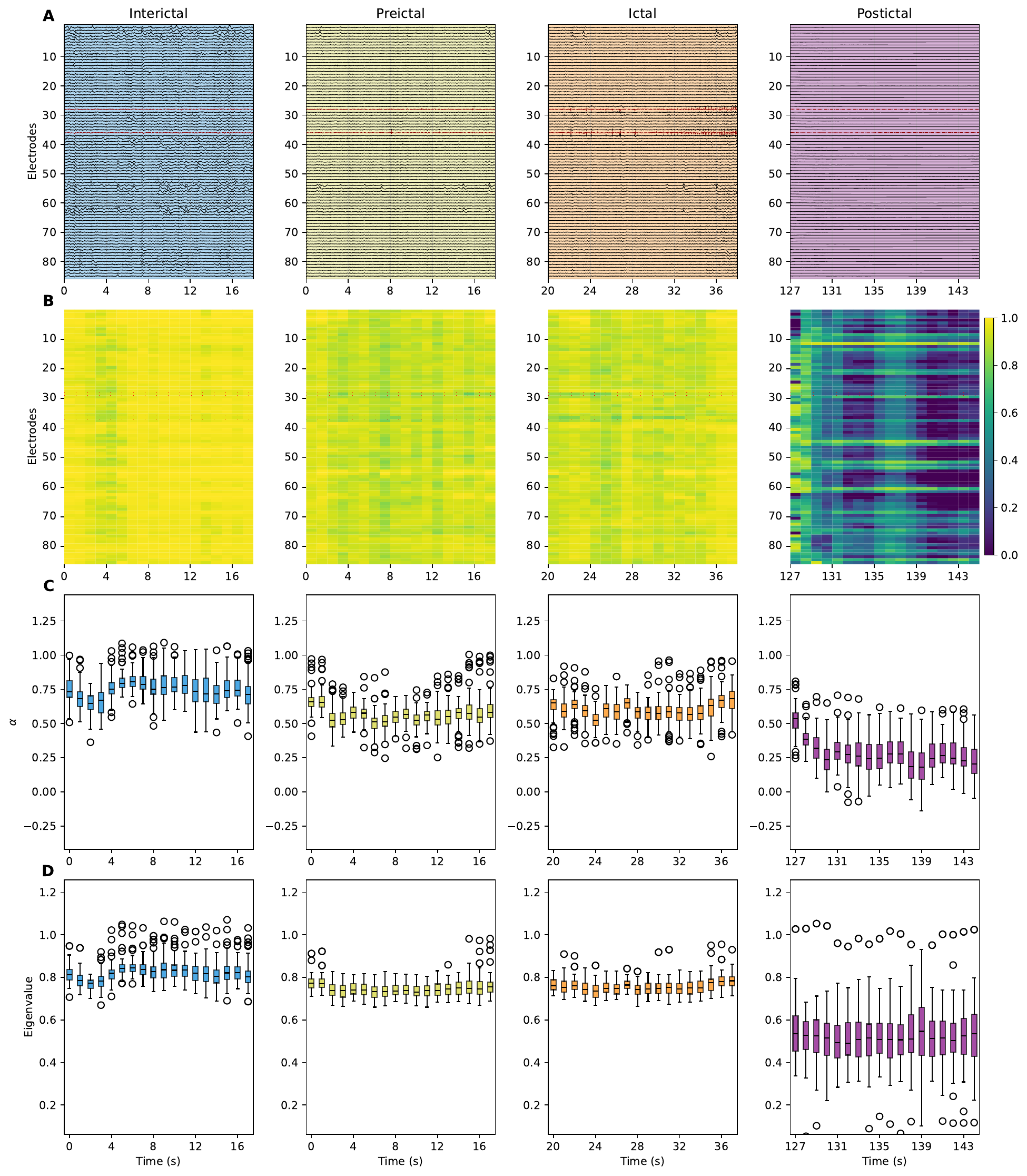}
    \caption{Representative visualization for Patient HUP64, Seizure 1. Each column corresponds to one segment (inter-ictal, pre-ictal, ictal, post-ictal). Rows show: (A) raw iEEG signals from 86 electrodes with seizure onset zone (SOZ) channels in red, (B) heatmap of FOS model R$^2$ fit values across electrodes, (C) boxplot of estimated $\alpha$ values across windows, and (D) boxplot of estimated eigenvalues across windows. Notably, the post-ictal segment did not have an accurate FOS fit, and both eigenvalues and $\alpha$ values are much lower than all other segments.}
    \label{fig:HUP64_4x4_block_1}
\end{figure}

\begin{figure}[h]
    \centering
    \includegraphics[width=\textwidth]{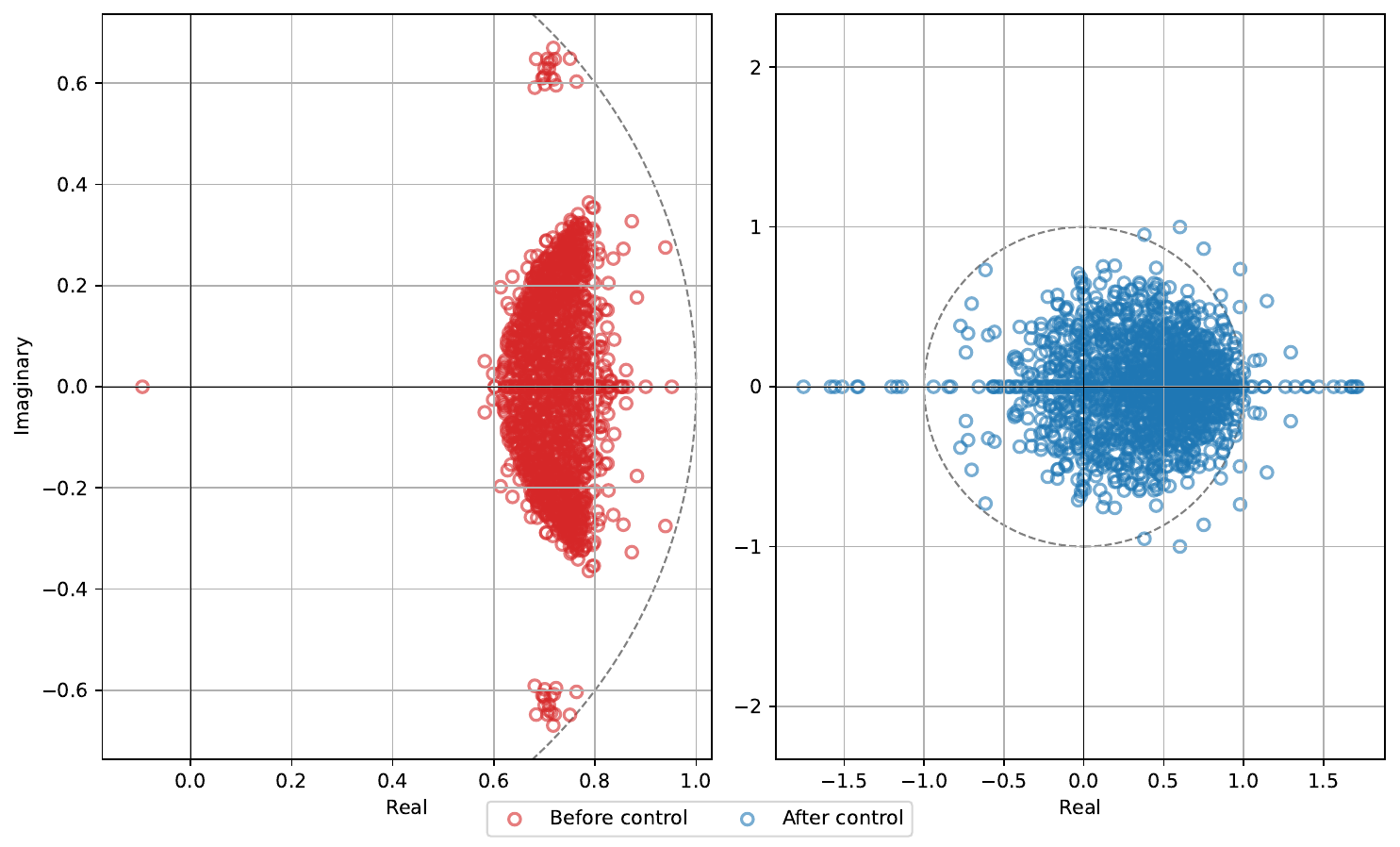}
    \caption{HUP78 Seizure 1. Red shows eigenvalues of the system matrix $A_0$ on the complex plane during seizure onset before applying stabilizing control. The fractional-order dynamical network is stable, according to our stability criterion (all eigenvalues within the unit circle). Blue shows eigenvalues of the simulated system matrix $A_0$ during seizure onset after applying stabilizing control. The control not only failed to improve stability but actually destabilized the system, shifting it from stable to unstable(max$|\lambda| \geq 1$), demonstrating a failure of the control algorithm.}
    \label{fig:HUP78_ctrl_block_1}
\end{figure}

\end{document}